\documentclass[11pt,twoside,letterpaper]{article} 
\usepackage{times,fancyhdr,graphicx}
\usepackage{url}
\def\ca{\rm{CA}\,{\small\rmfamily II}\,}
\def\snr{SN\,1993J~}
\def\duesnr{SN\, 2009ig~}
\def\velu{\rm\,{kms^{-1}}}
\def\2F1{~_2F_1}
\def\sun{\hbox{$\odot$}}
\def\aap{A\&A\,  }
\def\aj{AJ  }
\def\apj{ApJ\,  }
\def\apjl{ApJ\,  }
\def\apss{Astrophysics and Space Science  }

\def\mnras{MNRAS\,  }

\def\nat{Nature\,  }

  

\makeatletter
\def\cleardoublepage{\clearpage\if@twoside \ifodd\c@page\else%
    \hbox{}%
    \thispagestyle{empty}%
    \newpage%
    \if@twocolumn\hbox{}\newpage\fi\fi\fi}
\makeatother

\renewcommand{\thesection}{\arabic{section}.}
\renewcommand{\thesubsection}{\thesection\arabic{subsection}.}
\renewcommand{\thesubsubsection}{\thesubsection\arabic{subsubsection}.}

\def\figurename{Figure}
\makeatletter
\renewcommand{\fnum@figure}[1]{\figurename~\thefigure.}
\makeatother

\def\tablename{Table}
\makeatletter
\renewcommand{\fnum@table}[1]{\tablename~\thetable.}
\makeatother

\begin{document}
\title{
{\begin{flushleft} \vskip 0.45in
{\normalsize\bfseries\textit{Chapter~1}}
\end{flushleft}
\vskip 0.45in \bfseries\scshape  
Classical and relativistic laws   of
motion for spherical supernovas
}}
\author{\bfseries\itshape  
 Lorenzo Zaninetti \thanks{Email address: zaninetti@ph.unito.it}\\
Physics Department,
 via P.Giuria 1,\\ I-10125 Turin,Italy 
}

\date{}
\maketitle \thispagestyle{empty} \setcounter{page}{1}
\thispagestyle{fancy} \fancyhead{}
\fancyhead[L]{In: Book Title \\
Editor: Editor Name, pp. {\thepage-\pageref{lastpage-01}}} 
\fancyhead[R]{ISBN 0000000000  \\
\copyright~2007 Nova Science Publishers, Inc.} \fancyfoot{}
\renewcommand{\headrulewidth}{0pt}

\vspace{2in}

\vspace{.08in} \noindent \textbf{Keywords:} 
Supernovae
Supernova remnants    

\section*{Abstract}

We derive  some  first order differential
equations which model    the classical and the relativistic
thin layer approximations.
The circumstellar medium is
assumed to follow  a density profile 
which can be 
exponential,
Gaussian,
Plummer-like,
self-gravitating of Lane--Emden ($n=5$) type,
or power law.
The first order differential equations
are solved analytically,
or numerically, or by a series  expansion,
or by recursion,
or by Pad\'e approximation.
The initial conditions are chosen in order to model
the temporal evolution of SN 1993J over ten years.
The Pad\'e approximated  equations of motion  are applied to four 
SNRs: 
Tycho,
Cas A,   
Cygnus loop, 
and  SN~1006.

\section{Introduction}

The absorption features of supernovae (SN)
allow the determination of their expansion velocity,
$v$.
We select, among others, some results.
The  spectropolarimetry
(\ca IR triplet) of SN 2001el
gives a maximum velocity of $\approx 26000\velu$,
see  \cite{Wang2003}.
The same  triplet when searched in seven SN
of type Ia gives
$10400 \velu  \leq v \leq 17700 \velu $,
see Table I in \cite{Mazzali2005}.
A time series of eight  spectra in \duesnr
allows asserting that the velocity at the  \ca line,
for example,
decreases  in  12 days from 32000 $\velu$ to 21500$\velu$,
see Figure 9 in \cite{Marion2013}.

A recent analysis of 58 type Ia SN in Si II
gives  $9660 \velu  \leq v \leq 14820 \velu $,
see Table II
in \cite{Childress2014}.
Other examples for the maximum velocity of expansion 
are: 
$\approx 3700\, \velu$ , 
see Fig.~20.21 in \cite{Branch2017} and Fig.~6 
in \cite{Silverman2015}.

The  previous analysis  allow saying that
the maximum velocity so far observed for SN
is  $\frac{v}{c} \approx  0.123$, where
$c$ is the speed of light; this  observational fact
points to a relativistic equation of motion.

The temporal observations of SN such as \snr
establish a clear   relation
between the instantaneous radius of expansion $r$
and  the time $t$, of the
type  $r \propto t^{0.82}$,
see \cite{Marcaide2009},
and therefore  allows exploring variants of the thin layer
approximation.
The  previous observational facts
exclude  an SN propagation in a
circumstellar medium (CSM)
with constant density:
two  solutions of this type are
the  Sedov solution which,
scales  as  $r \propto t^{0.4}$,
 see \cite{Sedov1959,Dalgarno1987},
and the  momentum conservation  in  a thin layer
approximation, which
scales  as  $r \propto t^{0.25}$,
see \cite{Dyson1997,Padmanabhan_II_2001}.
Previous efforts  to  model
these observations
in the  framework
of the thin layer approximation
in a CSM governed by a power law,
 see  \cite{Zaninetti2011a},
or in the framework
in which the CSM   has a
constant density but swept mass regulated by a parameter called
porosity,
see \cite{Zaninetti2012c},
have been successfully explored.

An important feature of the various models
is based on the type of CSM  which surrounds the expansion.
As an example in the framework of classical shocks,
\cite{Chevalier1982a} and \cite{Chevalier1982b}
analyzed  self-similar solutions
with a CSM  of type $r^{-s}$, which means an
inverse power law dependence.
In the framework of the
Kompaneyets
equation, see \cite{Kompaneets1960},
for the motion of a shock wave in different plane-parallel
stratified media,
\cite{Olano2009}  considered four types of CSM.
It is therefore interesting to take into account
a self-gravitating CSM, which will give a physical basis to the
considered model.

The relativistic treatment  has been concentrated on the
determination of the Lorentz factor, $\gamma$,
for the ejecta in GRB; we report some  research in this regard:
\cite{Granot2006}  found  $30<; \gamma <;50$  for a
significant number  of GRBs,
\cite{Peer2007}  found $\gamma=305$ for GRB 970828  and
$\gamma=384$ for  GRB 990510,
\cite{Zou2010} found high values
for the sample of the GRBs considered
$30.5  \gamma <; 900$,
\cite{Aoi2010}
in the framework of a high-energy spectral
cutoff originating from the
creation of electron--positron pairs
found $\gamma \approx$ 600 for GRB 080916C,
\cite{Muccino2013}  found  $\gamma \approx 6.7 \times 10^2$
for GRB 090510.
The late stage of an SN is called supernova remnant (SNR)
and a lifetime of $\approx 1000 \,yr$ separates 
young SNRs from old SNRs.

In the present paper we
review the   standard
two-phase model for the expansion of an SN,
see Section \ref{standard},
and  five density profiles,
see Section \ref{secdensity}.
In  Section \ref{secclassic}
we derive the  differential equations
which model the thin layer approximation
for an SN in the presence of five types of medium.
A  relativistic treatment is carried out in
Section \ref{secrelativistic}.
The  application of the developed theory to \snr
is  split into the classical case,
see Section \ref{applicationsclassic},
and the relativistic case,
see Section  \ref{applicationsrelativistic}.
Section \ref{secastrosnr}
closes the derived equations of motion  to  four SNRs.

\section{The standard model}
\label{standard}

\begin{figure*}
\begin{center}
\includegraphics[width=7cm ]{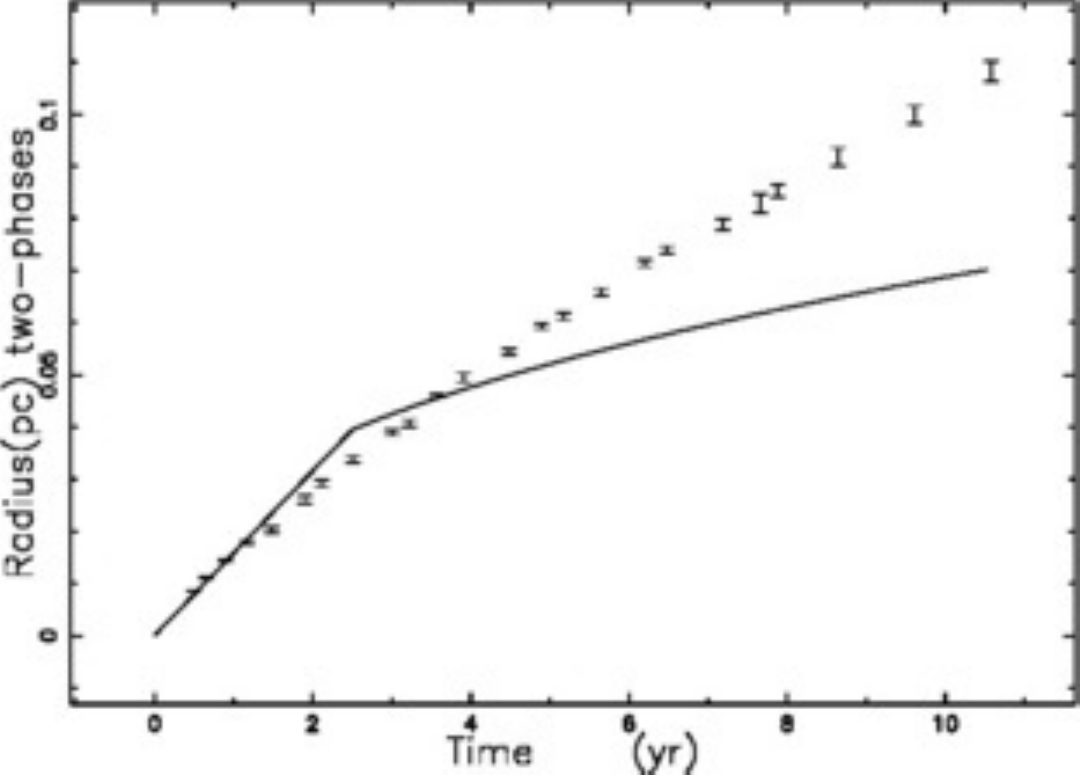}
\end {center}
\caption
{
Theoretical radius as given by the two-phase
solution (full line)
and astronomical data of \snr with
vertical error bars.
}
\label{1993duefasi}
    \end{figure*}

An SN  expands at a constant velocity until
the surrounding mass is
of the order of the solar mass.
The  time this takes, $t_M$,
is
\begin {equation}
t_M= 186.45\,{\frac {\sqrt [3]{{\it M_{\sun} }}}{\sqrt [3]{{\it n_0}}{\it
v_{10000}}}} \quad yr
\quad ,
\end{equation}
where $M_{\sun}$ is the number of solar masses
in the volume occupied by the SN,
$n_0$ is  the
number density  expressed  in particles~$\mathrm{cm}^{-3}$,
and $v_{10000}$ is the initial velocity expressed
in units of 10000\ km/s, see \cite{Dalgarno1987}.
A first law of motion for the $SN$
is the  Sedov  solution
\begin{equation}
R(t)=
 ({\frac {25}{4}}\,{\frac {{\it E}\,{t}^{2}}{\pi \,\rho}}  )^{1/5}
\quad ,
\label{sedov}
\end{equation}
where $E$ is the energy injected into the process
and   $t$ is the  time,
see~\cite{Sedov1959,Dalgarno1987}.
Our astrophysical  units are: time, ($t_1$), which
is expressed  in years;
$E_{51}$, the  energy in  $10^{51}$ erg;
$n_0$,  the
number density  expressed  in particles~$\mathrm{cm}^{-3}$~
(density~$\rho=n_0m$, where $m = 1.4m_{\mathrm {H}}$).
In these units, equation ~(\ref{sedov}) becomes
\begin{equation}
R(t) \approx  0.313\,\sqrt [5]{{\frac {{\it E_{51}}\,{{\it t_1}}^{2}}{{\it n_0}}}
}~{pc}
\quad .
\label{sedovastro}
\end{equation}
The Sedov solution scales as $t^{0.4}$.
We are now ready to couple
the Sedov phase with the free expansion phase
\begin{equation}
 R(t)  = \{ \begin{array}{ll}
0.0157t\,\mbox{pc} &
\mbox {if $t \leq 2.5$ yr } \\
0.0273\,\sqrt [5]t^2\, \mbox{pc}   &
\mbox {if $t >;    2.5$ yr.}
            \end{array}
\label{twophases}
\nonumber
\end{equation}
This two-phase solution is obtained with
the following parameters
$M_{\sun}$ =1 ,
$n_0 =1.127 \times10^5$,
$E_{51}=0.567$
and Figure~\ref{1993duefasi} presents its  temporal behavior
as well as the data.

A similar  model  is reported in \cite{Spitzer1978} with the
difference that the first phase ends  at $t=60$\ yr against our
$t=2.5$\ yr.
A careful  analysis of Figure~\ref{1993duefasi}   reveals that
the standard two-phase model does not fit
the observed radius--time relation  for  \snr.

\section{Density profiles for the CSM}

\label{secdensity}

This  section  introduces five  density profiles
for the CSM:
an exponential profile,
a  Gaussian    profile,
a Plummer-like profile,
a self-gravitating profile  of Lane--Emden type,
and   a power law profile.

\label{secmomentum}

\subsection{The exponential profile}

This density  is  assumed to have the following
exponential dependence on $r$
in spherical  coordinates:
\begin{equation}
 \rho(r;r_0,b,\rho_0) =
\rho_0  \exp{(-\frac{(r-r_0)}{b})}
\quad ,
\label{profexponential}
\end{equation}
where $b$ represents the scale.
The piece-wise  density is
\begin{equation}
 \rho (r;r_0,b,\rho_0)  = \left\{ \begin{array}{ll} 
            \rho_0                      & \mbox         {if $r \leq r_0 $ } \\
            \rho_0 \exp{-(\frac{(r-r_0)}{b})}   & \mbox {if $r >;    r_0 $ } 
            \end{array}
            \right.
\label{profile_exponential}
\end{equation}
The total mass swept,   $M(r;r_0,b,\rho_0) $,
in the interval $[0,r]$ is
\begin{eqnarray}
M(r;r_0,b,\rho_0) = \nonumber  \\
\frac{4}{3}\,\rho_{{0}}\pi\,{r_{{0}}}^{3}
\nonumber \\
-4\,b \left( 2\,{b}^{2}+2\,br+{r}^{2}
 \right) \rho_{{0}}{{\rm e}^{{\frac {r_{{0}}-r}{b}}}}\pi+4\,b \left( 2
\,{b}^{2}+2\,br_{{0}}+{r_{{0}}}^{2} \right) \rho_{{0}}\pi
\quad .
\end{eqnarray}

\subsection{The Gaussian  profile}

This density  has  the 
Gaussian  dependence 
\begin{equation}
 \rho(r;r_0,b,\rho_0) =
\rho_0  \exp{(-\frac{1}{2}\frac{r^2}{b^2})}
\quad ,
\label{prof_gaussian}
\end{equation}
and the piece-wise  density is
\begin{equation}
 \rho (r;r_0,b,\rho_0)  = \left\{ \begin{array}{ll} 
            \rho_0                      & \mbox         {if $r \leq r_0 $ } \\
            \rho_0\exp{(-\frac{1}{2}\frac{r^2}{b^2})}   & \mbox {if $r >;    r_0 $ } 
            \end{array}
            \right.
\end{equation}

The total mass swept,   $M(r;r_0,b,\rho_0) $,
in the interval $[0,r]$ is
\begin{eqnarray}
M(r;r_0,b,\rho_0) = \nonumber  \\
\frac{4}{3}\,\rho_{{0}}\pi\,{r_{{0}}}^{3}+4\,\rho_{{0}}\pi\, \big  ( -{{\rm e}^
{-\frac{1}{2}\,{\frac {{r}^{2}}{{b}^{2}}}}}r{b}^{2}+\frac{1}{2}\,{b}^{3}\sqrt {\pi}
\sqrt {2}{\rm erf}   (\frac{1}{2}\,{\frac {\sqrt {2}r}{b}}  )  \big )\nonumber \\
 -
4\,\rho_{{0}}\pi\, \big  ( -{{\rm e}^{-\frac{1}{2}\,{\frac {{r_{{0}}}^{2}}{{b}^
{2}}}}}r_{{0}}{b}^{2}+\frac{1}{2}\,{b}^{3}\sqrt {\pi}\sqrt {2}{\rm erf}   (
\frac{1}{2}\,{\frac {\sqrt {2}r_{{0}}}{b}}  ) \big  ) 
\quad ,
\end{eqnarray}
where ${\rm erf}$ is the error function,    
see \cite{NIST2010}.

\subsection{The Plummer profile}

The Plummer-like  density profile, after \cite{Plummer1911},
is
\begin{equation}
\rho(r;R_{flat}) =
\rho_c  ({\frac {R_{flat}}{({R_{flat}^2 +r^2})^{1/2}}}  )^{\eta}
\label{densitaplummerflat}
\nonumber
\quad
\end{equation}
where $r$ is the distance from the center,
$\rho$    is the density,
$\rho_c$  is the density at the center,
$R_{flat}$ is the distance up to which the density is nearly constant,
and $\eta$  is the power law exponent at large values of $r$,
see \cite{Whitworth2001} for more details.
The following transformation, $R_{flat}= \sqrt{3} b $,
gives the  Plummer-like profile, which can be compared
with the Lane--Emden profile
\begin{equation}
\rho(r;b) =
\rho_c \bigl ( \frac{1}{1+ \frac{1}{3}\,{\frac {{r}^{2}}{{b}^{2}}}}
 \bigr ) ^{\eta/2}
\quad .
\label{densitaplummer}
\quad
\end{equation}
At low values of $r$,  the Taylor expansion of the Plummer-like profile
can be taken:
 \begin{equation}
\rho(r;b) \approx \rho_{{c}}( 1-1/6\,{\frac {\eta\,{r}^{2}}{{b}^{2}}})
\label{scalingplummerlow}
\nonumber
\quad ,
\end{equation}
and at  high  values of $r$,  the
behavior  of the Plummer-like profile is
\begin{equation}
\rho(r;b) \sim \rho_c
( \sqrt{3}\,b ) ^{\eta} ( \frac{1}{r} ) ^{
\eta}
\nonumber
\label{scalingplummerhigh}
\quad .
\end{equation}

The  total  mass $M(r;b)$ contained between
0 and  $r$
is
\begin{equation}
M(r;b) = \int_0^r   4 \pi r^2   \rho(r;b) dr
= \frac{PN}{PD}
\quad ,
\end{equation}
where
\begin{eqnarray}
PN =-\sqrt {3}b\rho_{{c}}\pi \, ( 4\,
{\mbox{$_2$F$_1$}(\eta/2,-3/2+\eta/2;\,-1/2+\eta/2;\,-3\,{\frac {{b}^{2}}{{r}^{2}}})} \times  \nonumber \\
\times \Gamma  ( -\eta/2+5/2 ) \Gamma  ( \eta/2 ) \cos
 ( 1/2\,\pi \,\eta ) {r}^{3-\eta}{3}^{1/2+\eta/2}{b}^{\eta-
1} \nonumber \\
-9\,{\pi }^{3/2}{b}^{2}\eta+27\,{\pi }^{3/2}{b}^{2} )
\nonumber
\end{eqnarray}
and
\begin{equation}
PD=3\,\cos ( 1/2\,\pi \,\eta ) \Gamma  ( -\eta/2+5/2
 )  ( \eta-3 ) \Gamma  ( \eta/2 )
\nonumber
\quad ,
\end{equation}
where ${\2F1(a,b;\,c;\,z)}$ is the
regularized hypergeometric
function
\cite{Abramowitz1965,NIST2010}.
The above expression simplifies when $\eta=6$ , $M(r;b)_6$,
\begin{eqnarray}
M(r;b)_6 =
{\frac {27\,\rho_{{c}}\pi \,{b}^{7}\sqrt {3}}{2\, ( 3\,{b}^{2}+{r
}^{2} ) ^{2}}\arctan ( 1/3\,{\frac {r\sqrt {3}}{b}}
 ) }+9\,{\frac {\rho_{{c}}\pi \,{b}^{5}\sqrt {3}{r}^{2}}{
 ( 3\,{b}^{2}+{r}^{2} ) ^{2}}\arctan ( 1/3\,{\frac {r
\sqrt {3}}{b}} ) }+ \nonumber \\
3/2\,{\frac {\rho_{{c}}\pi \,{b}^{3}\sqrt {3}
{r}^{4}}{ ( 3\,{b}^{2}+{r}^{2} ) ^{2}}\arctan ( 1/3\,{
\frac {r\sqrt {3}}{b}} ) }-{\frac {27\,\rho_{{c}}\pi \,{b}^{6}r
}{2\, ( 3\,{b}^{2}+{r}^{2} ) ^{2}}}+9/2\,{\frac {\rho_{{c}}
\pi \,{b}^{4}{r}^{3}}{ ( 3\,{b}^{2}+{r}^{2} ) ^{2}}}
 \quad .
 \nonumber
\end{eqnarray}
The astrophysical version of the total mass is
\begin{equation}
M(r_{pc};b_{pc}) = \frac{PNA}{PDA}  \,M_{\sun}
\quad ,
\nonumber
\end{equation}
with
\begin{eqnarray}
PNA= - 2.47\,10^{-10}\,{b_{{{\it pc}}}}^{3}n_{{0}} [
 1.02\,10^{10}\,\arctan (  1.73\,{\frac {b_{{{\it pc}}}}{r
_{{{\it pc}}}}}  ) {b_{{{\it pc}}}}^{4}
 \nonumber \\
+ 6.8\,10^9\,\arctan
 (  1.73\,{\frac {b_{{{\it pc}}}}{r_{{{\it pc}}}}}
  ) {b_{{{\it pc}}}}^{2}{r_{{{\it pc}}}}^{2}
  + 1.13\,10^9\,
\arctan (  1.73\,{\frac {b_{{{\it pc}}}}{r_{{{\it pc}}}}}
  ) {r_{{{\it pc}}}}^{4}- 1.6\,10^{10}\,{b_{{{\it pc}}}}^{4}
\nonumber \\
  +  5.89\,10^9\,{b_{{{\it pc}}}}^{3}r_{{{\it pc}}}- 1.06\,10^{10}\,{b_{
{{\it pc}}}}^{2}{r_{{{\it pc}}}}^{2}- 1.96\, 10^9\,b_{{{\it pc}}}{r_{
{{\it pc}}}}^{3}- 1.78\,10^9\,{r_{{{\it pc}}}}^{4}  ]
\nonumber
\end{eqnarray}
and
\begin{equation}
PDA=\left(  3.0\,{b_{{{\it pc}}}}^{2}+{r_{{{\it pc}}}}^{2} \right) ^{2}
\quad ,
\nonumber
\end{equation}
where ${b_{{{\it pc}}}}$ is $b$ expressed in pc,
${r_{{{\it pc}}}}$ is $r$ expressed in pc,
and $n_0$ is the same as in
equation~(\ref{sedovastro}).
More details can be found in \cite{Zaninetti2014f}.

\subsection{The Lane--Emden profile}

The self gravitating sphere of polytropic
gas is governed
by the Lane--Emden differential equation
of the second order
\begin{equation}
{\frac {d^{2}}{d{x}^{2}}}Y ( x ) +2\,{\frac {{\frac {d}{dx}
}Y ( x ) }{x}}+ ( Y ( x )  ) ^{n}=0
\quad ,
\nonumber
\end{equation}
where $n$  is an integer, see
\cite{Lane1870,Emden1907,Chandrasekhar_1967,Binney2011,Zwillinger1989}.

The solution  $Y ( x )_n$
has  the density profile
\begin{equation}
\rho = \rho_c Y ( x )_n^n
\quad ,
\nonumber
\end{equation}
where $\rho_c$ is the density at $x=0$.
The pressure $P$ and temperature $T$ scale as
\begin{equation}
P = K \rho^{1 +\frac{1}{n}}
\quad ,
\label{pressure}
\end{equation}
\begin{equation}
 T = K^{\prime} Y(x)
\label{temperature}
\quad ,
\end{equation}
where  $K$ and $K^{\prime}$   are two  constants,
for more details, see \cite{Hansen1994}.

Analytical solutions exist for $n=0$, 1 and 5;
that  for $n$=0 is
\begin{equation}
Y(x) = \frac{sin(x)}{x}
\quad,
\nonumber
\end{equation}
and has therefore an oscillatory behavior.
The analytical  solution for $n$=5 is
\begin{equation}
Y(x) ={\frac {1}{{(1+ \frac{{x}^{2}}{3})^{1/2}}} }
\quad ,
\nonumber
\end{equation}
and the density for $n$=5 is
\begin{equation}
\rho(x) =\rho_c {\frac {1}{{(1+ \frac{{x}^{2}}{3})^{5/2}}} }
\label{densita5}
\quad .
\end{equation}

The   variable  $x$   is   non-dimensional
and  we now  introduce the
new variable $x=r/b$
\begin{equation}
\rho(r;b) =\rho_c {\frac {1}{{(1+ \frac{{r}^{2}}{3b^2})^{5/2}}} }
\label{densita5b}
\label{profile_lane}
\nonumber
\quad .
\end{equation}
This profile is a particular case, $\eta=5$,
of the Plummer-like profile as given by equation~(\ref{densitaplummer}).
At low values of $r$,  the Taylor expansion of this profile is
 \begin{equation}
\rho(r;b) \approx \rho_{{c}} ( 1-5/6\,{\frac {{r}^{2}}{{b}^{2}}} )
\nonumber
\label{scalinglow}
\quad ,
\end{equation}
and at  high  values of $r$,  its
 behavior is
\begin{equation}
\rho(r;b) \sim 9\,{\frac {\rho_{{c}}\sqrt {3}{b}^{5}}{{r}^{5}}}
\label{scalinghigh}
\quad .
\end{equation}

The  total mass $M(r;b)$ contained between
0 and  $r$
is
\begin{equation}
M(r;b) = \int_0^r   4 \pi r^2   \rho(r;b) dr
=\frac
{
4\,{b}^{3}{r}^{3}\rho_c\,\pi \,\sqrt {3}
}
{
( 3\,{b}^{2}+{r}^{2} ) ^{3/2}
}
\quad ,
\label{integralemassa}
\end{equation}
or  in solar units
\begin{equation}
M(r_{pc};b_{pc}) =
\frac
{
{ 2.2\times 10^{55}}\,{b_{{{\it pc}}}}^{3}{r_{{{\it pc}}}}^{3}
n_{{0}}
}
{
\left( { 2.85\times 10^{37}}\,{b_{{{\it pc}}}}^{2}+{
 9.52\times 10^{36}}\,{r_{{{\it pc}}}}^{2} \right) ^{3/2}
}
 \,M_{\sun}
\quad .
\nonumber
\end{equation}
The total mass of the profile  can be found
calculating the limit as $r \to \infty$
of equation~(\ref{integralemassa})
\begin{equation}
M(\infty;b) = \lim_{r\to \infty}M(r;b) = 4\,{b}^{3}{\it \rho_c}\,\pi \,\sqrt
{3}
\label{totalmass}
\quad .
\end{equation}

\subsection{The power law profile}
\label{powerlawsubsec}

We now  assume that the CSM
around the SN
scales with the following piecewise dependence
(which avoids a pole at $r=0$)
\begin{equation}
 \rho (r;r_0,d)  = \{ \begin{array}{ll}
            \rho_c                      & \mbox {if $r \leq r_0 $ } \\
            \rho_c (\frac{r_0}{r})^d    & \mbox {if $r >;    r_0 $.}
            \end{array}
\label{piecewise}
\end{equation}

The mass swept, $M_0$,
in the interval [0,$r_0$]
is
\begin{equation}
M_0 =
\frac{4}{3}\,\rho_{{0}}\pi \,{r_{{0}}}^{3}
\quad .
\nonumber
\end{equation}
The total mass swept, $ M(r;r_0,d) $,
in the interval [0,r]
is
\begin{eqnarray}
M (r;r_0,d)=
-4\,{r}^{3}\rho_{{c}}\pi \, ( {\frac {r_{{0}}}{r}} ) ^{d}
 (d -3 ) ^{-1}   \nonumber \\
+4\,{\frac {\rho_{{c}}\pi \,{r_{{0}}}^{3}}{d-
3}}
+ \frac{4}{3}\,\rho_{{c}}\pi \,{r_{{0}}}^{3}
\quad .
\nonumber
\label{masspowerlaw}
\end{eqnarray}
or  in solar units
\begin{equation}
M(r_{pc};{r_{{0,{\it pc}}}},d) =
\frac
{
 3.14\,n_{{0}} \left(  0.137\,{r_{{{\it pc}}}}^{3}
 \left( {\frac {r_{{0,{\it pc}}}}{r_{{{\it pc}}}}} \right) ^{d}-
 0.0459\,{r_{{0,{\it pc}}}}^{3}d \right)
}
{
3-d
}
\,M_{\sun}
\quad ,
\nonumber
\end{equation}
where  ${r_{{0,{\it pc}}}}$ is  $r_0$ expressed in pc.

\section{Classical conservation of momentum}

\label{secclassic}

This section reviews  the standard
equation of motion in the case of
the thin layer approximation in the presence of 
a CSM with constant density
and derives the equation of motion  under the conditions
of each of the five density profiles for the  density of the CSM.

\subsection{Motion with constant density}

In the case of a constant  density of the CSM,
$\rho_c$, the differential equation
which models momentum conservation
is
\begin{equation}
\frac{4}{3}\,\pi \, ( r ( t )  ) ^{3}\rho_{{c}}{\frac {d
}{dt}}r ( t ) -\frac{4}{3}\,\pi \,{{\it r_0}}^{3}\rho_{{c}}v_{{0}}=0
\quad ,
\nonumber
\end{equation}
where the initial  conditions
are  $r=r_0$  and   $v=v_0$
when $t=t_0$.
The variables can be separated and
the radius as a function of the time
is
\begin{equation}
r(t)= \sqrt [4]{4\,{r_{{0}}}^{3}v_{{0}} ( t-t_{{0}} ) +{r_{{0}}}^
{4}}
\quad ,
\nonumber
\end{equation}
and its   behavior
as $\quad t  \rightarrow \infty$  is
\begin{eqnarray}
r(t) =\sqrt {2}{r_{{0}}}^{3/4}\sqrt [4]{v_{{0}}}\sqrt [4]{t-t_{{0}}}+ \frac{1}{16}\,{
\frac {\sqrt {2}{r_{{0}}}^{7/4}}{{v_{{0}}}^{3/4} ( t-t_{{0}}
 ) ^{3/4}}}.
\nonumber
\end{eqnarray}
The velocity  as a function of time is
\begin{equation}
v(t)=
\frac
{
{r_{{0}}}^{3}v_{{0}}
}
{
 ( 4\,{r_{{0}}}^{3}v_{{0}} ( t-t_{{0}} ) +{r_{{0}}}^{4
} ) ^{3/4}
}
\quad .
\nonumber
\end{equation}

\subsection{Motion with exponential profile}

Assuming an exponential profile as given by 
equation~(\ref{profile_exponential}),
the velocity is
\begin{equation}
\frac{dr}{dt} = \frac{NE}{DE}
\quad ,
\label{vel_exp}
\end{equation}
where
\begin{eqnarray}
NE =
{-{r_{{0}}}^{3}v_{{0}}}
\nonumber  
\quad,
\end{eqnarray}
and  
\begin{eqnarray}
DE=
6\,{{\rm e}^{{\frac {r_{{0}}-r}{b}}}}{b}^{3}+6\,{{\rm e}^{{\frac {r_{{0
}}-r}{b}}}}{b}^{2}r
\nonumber \\
+3\,{{\rm e}^{{\frac {r_{{0}}-r}{b}}}}b{r}^{2}-{r_{
{0}}}^{3}-3\,{r_{{0}}}^{2}b-6\,r_{{0}}{b}^{2}-6\,{b}^{3}
\quad .
\nonumber
\end{eqnarray}
In the above differential equation of the first order in $r$, the
variables can be separated and integration
gives the following nonlinear equation:
\begin{eqnarray}
\frac {1}{{r_{{0}}}^{3}{\it v_0}}
\bigg (
18\,{{\rm e}^{{\frac {r_{{0}}-r}{b}}}}{b}^{4}+12\,{{\rm e}^{{\frac {r_
{{0}}-r}{b}}}}{b}^{3}r+3\,{{\rm e}^{{\frac {r_{{0}}-r}{b}}}}{b}^{2}{r}
^{2}-{r_{{0}}}^{4}-3\,{r_{{0}}}^{3}b
\nonumber\\
+{r_{{0}}}^{3}r-9\,{r_{{0}}}^{2}{b
}^{2}+3\,{r_{{0}}}^{2}br-18\,{b}^{3}r_{{0}}+6\,r_{{0}}{b}^{2}r-18\,{b}
^{4}+6\,{b}^{3}r
\bigg )
\nonumber \\
=\left( t-{\it t_0} \right) 
\label .
\label{eqn_nl_exp}
\end{eqnarray}
In this case is not possible to find an analytical  
solution for the radius, $r$,
as a function of  time.
We therefore apply 
the Pad\'e rational polynomial 
approximation of degree 2 in the numerator
and degree 1 in the denominator about the point $r=r_0$ 
to the  left-hand  side of 
equation~(\ref{eqn_nl_exp}):
\begin{equation}
\frac
{
- \left( r_{{0}}-r \right)  \left( -5\,br-br_{{0}}-2\,rr_{{0}}+2\,{r_{
{0}}}^{2} \right) 
}
{
2\,v_{{0}} \left( 2\,br-5\,br_{{0}}-rr_{{0}}+{r_{{0}}}^{2} \right)  
}
= t-t_0 
\quad .
\end{equation}
The resulting Pad\'e  approximant  for the radius $r_{2,1}$ is
\begin{eqnarray}
r_{2,1}=\frac{1}{2\,r_{{0}}+5\,b}
\Bigg  ( r_{{0}}tv_{{0}}-r_{{0}}{\it t_0}\,v_{{0}}-2\,btv_{{0}}+2\,b{\it t_0}\,v_
{{0}}+2\,{r_{{0}}}^{2}+2\,r_{{0}}b
\nonumber \\
+ \biggl ( {4\,{b}^{2}{t}^{2}{v_{{0}}}^{
2}-8\,{b}^{2}t{\it t_0}\,{v_{{0}}}^{2}+4\,{b}^{2}{{\it t_0}}^{2}{v_{{0}}
}^{2}-4\,b{t}^{2}r_{{0}}{v_{{0}}}^{2}}
\nonumber\\
{
+8\,bt{\it t_0}\,r_{{0}}{v_{{0}}}^
{2}-4\,b{{\it t_0}}^{2}r_{{0}}{v_{{0}}}^{2}+{t}^{2}{r_{{0}}}^{2}{v_{{0}
}}^{2}-2\,t{\it t_0}\,{r_{{0}}}^{2}{v_{{0}}}^{2}+{{\it t_0}}^{2}{r_{{0}}
}^{2}{v_{{0}}}^{2}
}
\nonumber \\
{
+42\,{b}^{2}tr_{{0}}v_{{0}}
-42\,{b}^{2}{\it t_0}\,r_{
{0}}v_{{0}}+6\,bt{r_{{0}}}^{2}v_{{0}}-6\,b{\it t_0}\,{r_{{0}}}^{2}v_{{0
}}+9\,{r_{{0}}}^{2}{b}^{2}} \biggl )^{\frac{1}{2}} 
\Bigg )
\quad ,
\label{rmotionexp} 
\end{eqnarray}
and the velocity is
\begin{equation}
v_{2,1}=\frac{dr_{2,1}}{dt} =\frac{NVE}{DVE}
\quad ,
\label{vmotionexp} 
\end{equation}
\begin{eqnarray} 
NVE =
4\,v_{{0}} \Big  \{    ( -b/2+1/4\,r_{{0}}   )\times  \nonumber \\
\sqrt {4\,   ( 
b-\frac{1}{2}\,r_{{0}}   ) ^{2}   ( t-t_{{0}}   ) ^{2}{v_{{0}}}^{2}
+42\,   ( b+1/7\,r_{{0}}   )    ( t-t_{{0}}   ) br_{{0}}
v_{{0}}+9\,{r_{{0}}}^{2}{b}^{2}}+\nonumber  \\  ( 3/4\,b+   ( t/4-1/4\,t_{{0
}}   ) v_{{0}}   ) {r_{{0}}}^{2}+{\frac {21\,r_{{0}}b}{4}
   ( v_{{0}}   ( -{\frac {4\,t}{21}}+{\frac {4\,t_{{0}}}{21}}
   ) +b   ) }
\nonumber  \\
+{b}^{2}v_{{0}}   ( t-t_{{0}}   ) 
  \Big \}
\quad ,
\end{eqnarray}
and
\begin{eqnarray}
DVE =   \nonumber \\
\sqrt {4\,   ( b-\frac{1}{2}\,r_{{0}}   ) ^{2}   ( t-t_{{0}}
   ) ^{2}{v_{{0}}}^{2}+42\,   ( b+1/7\,r_{{0}}   )    ( 
t-t_{{0}}   ) br_{{0}}v_{{0}}+9\,{r_{{0}}}^{2}{b}^{2}} \times
\nonumber \\
  ( 2\,r
_{{0}}+5\,b   ) 
\quad .
\end{eqnarray}

\subsection{Motion with Gaussian profile}

Assuming a   Gaussian  profile as given by 
equation~(\ref{prof_gaussian}) 
the velocity is
\begin{equation}
\frac{dr}{dt} = \frac{NG}{DG}
\quad ,
\label{vel_gaussian}
\end{equation}
where
\begin{eqnarray}
NG= -2\,{r_{{0}}}^{3}v_{{0}}
\end{eqnarray}
and  
\begin{eqnarray}
DG= 
-3\,{b}^{3}\sqrt {\pi}\sqrt {2}{\rm erf} \left(\frac{1}{2}\,{\frac {\sqrt {2}r
}{b}}\right)
\nonumber \\
+3\,{b}^{3}\sqrt {\pi}\sqrt {2}{\rm erf} \left(\frac{1}{2}\,{
\frac {\sqrt {2}r_{{0}}}{b}}\right)+6\,{{\rm e}^{-\frac{1}{2}\,{\frac {{r}^{2}
}{{b}^{2}}}}}r{b}^{2}
\nonumber \\
-6\,{{\rm e}^{-\frac{1}{2}\,{\frac {{r_{{0}}}^{2}}{{b}^{2
}}}}}r_{{0}}{b}^{2}-2\,{r_{{0}}}^{3}
\quad .
\end{eqnarray}
The appropriate  nonlinear equation  is
\begin{eqnarray}
\frac{1}{2\,{r_{{0}}}^{3}v_{{0}}} 
\bigg (
  ( -12\,{b}^{4}+6\,r_{{0}}   ( r-r_{{0}}   ) {b}^{2}
   ) {{\rm e}^{-\frac{1}{2}\,{\frac {{r_{{0}}}^{2}}{{b}^{2}}}}}+12\,{b}^{4
}{{\rm e}^{-\frac{1}{2}\,{\frac {{r}^{2}}{{b}^{2}}}}}
\nonumber \\
-3\,\sqrt {\pi}{\rm erf} 
  (\frac{1}{2}\,{\frac {\sqrt {2}r_{{0}}}{b}}  )\sqrt {2}{b}^{3}r+3\,{b
}^{3}\sqrt {\pi}\sqrt {2}{\rm erf}   (\frac{1}{2}\,{\frac {\sqrt {2}r}{b}}
  )r
\nonumber \\
+2\,{r_{{0}}}^{3}   ( r-r_{{0}}   )
\bigg )
= t-t_0 \, . 
\label{eqn_nl_gaussian}
\end{eqnarray}
The Pad\'e rational polynomial 
approximation of degree 2 in the numerator
and degree 1 in the denominator 
about $r=r_0$ 
for the left-hand side of the above equation gives
\begin{eqnarray}
\frac
{
1
}
{
2\,v_{{0}} \left( 2\,{b}^{2}r-5\,r_{{0}}{b}^{2}-r{r_{{0}}}^{2}+{r_{{0}
}}^{3} \right)
}
\Bigg (
-   ( r-r_{{0}}   )   \bigg ( 9\,{{\rm e}^{-\frac{1}{2}\,{\frac {{r_{{0}}
}^{2}}{{b}^{2}}}}}{b}^{2}r
\nonumber \\
-9\,{{\rm e}^{-\frac{1}{2}\,{\frac {{r_{{0}}}^{2}}{{
b}^{2}}}}}r_{{0}}{b}^{2}-4\,{b}^{2}r+10\,r_{{0}}{b}^{2}+2\,r{r_{{0}}}^
{2}-2\,{r_{{0}}}^{3} \bigg  )
\Bigg )
= t-t_0 \, .
\end{eqnarray}
The resulting Pad\'e  approximant  for the radius $r_{2,1}$ is
\begin{eqnarray}
r_{2,1}=  
\frac{1}
{
9\,{{\rm e}^{-\frac{1}{2}\,{\frac {{r_{{0}}}^{2}}{{b}^{2}}}}}{b}^{2}+2\,{r_{{0
}}}^{2}-4\,{b}^{2}
}
\Bigg \{
9\,{{\rm e}^{-\frac{1}{2}\,{\frac {{r_{{0}}}^{2}}{{b}^{2}}}}}r_{{0}}{b}^{2}-2
\,{b}^{2}tv_{{0}}
\nonumber \\
+2\,{b}^{2}{\it t_0}\,v_{{0}}+{r_{{0}}}^{2}tv_{{0}}-{r
_{{0}}}^{2}{\it t_0}\,v_{{0}}-7\,r_{{0}}{b}^{2}+2\,{r_{{0}}}^{3}
\nonumber \\
+ \bigg [ 
{54\,{b}^{4}r_{{0}}v_{{0}} \Big( t-{\it t_0} \Big) {{\rm e}^{-\frac{1}{2}\,{
\frac {{r_{{0}}}^{2}}{{b}^{2}}}}}
}
\nonumber \\
{
+4\, \Big(  \Big( t-{\it t_0}
 \Big)  \Big( {b}^{2}-\frac{1}{2}\,{r_{{0}}}^{2} \Big) v_{{0}}-\frac{3}{2}\,r_{{0
}}{b}^{2} \Big) ^{2}}
\bigg ]^{\frac{1}{2}}
\Bigg \}  
\label{rmotiongauss}
\, , 
\end{eqnarray}
and the velocity is
\begin{equation}
v_{2,1}=\frac{dr_{2,1}}{dt} =\frac{NVG}{DVG}
\quad ,
\label{vmotiongauss} 
\end{equation}
\begin{eqnarray} 
NVG =
-   \Bigg ( -27\,{{\rm e}^{-\frac{1}{2}\,{\frac {{r_{{0}}}^{2}}{{b}^{2}}}}}r_{{0}
}{b}^{4}+   ( 2\,{b}^{2}-{r_{{0}}}^{2}   )    ( v_{{0}}
   ( t-t_{{0}}   ) {r_{{0}}}^{2}
\nonumber \\
+3\,r_{{0}}{b}^{2}
-2\,v_{{0}}{b
}^{2}   ( t-t_{{0}}   ) 
+ \bigg \{ {54\,{b}^{4}r_{{0}}v_{{0}}
   ( t-t_{{0}}   ) {{\rm e}^{-\frac{1}{2}\,{\frac {{r_{{0}}}^{2}}{{b}^{
2}}}}}}
\nonumber \\
{ 
+4\,   (    ( t-t_{{0}}   )    ( {b}^{2}-\frac{1}{2}\,{r_{{0
}}}^{2}   ) v_{{0}}-3/2\,r_{{0}}{b}^{2}   ) ^{2}}   ) 
  \bigg \} ^{\frac{1}{2}} \Bigg ) v_{{0}}
\quad ,
\end{eqnarray}
and
\begin{eqnarray} 
DVG =
\Bigg \{ {54\,{b}^{4}r_{{0}}v_{{0}}   ( t-t_{{0}}   ) {{\rm e}^{-1
/2\,{\frac {{r_{{0}}}^{2}}{{b}^{2}}}}}+4\,   (    ( t-t_{{0}}
   )    ( {b}^{2}-\frac{1}{2}\,{r_{{0}}}^{2}   ) v_{{0}}
}
\nonumber \\
{
-3/2\,r_{{0
}}{b}^{2}   ) ^{2}}   
\Bigg \}^{\frac{1}{2}}
( 9\,{{\rm e}^{-\frac{1}{2}\,{\frac {{r_{{0}}}^{2
}}{{b}^{2}}}}}{b}^{2}+2\,{r_{{0}}}^{2}-4\,{b}^{2}   ) 
\quad .
\end{eqnarray}

\subsection{Motion with Plummer profile}

The case of a Plummer-like profile  for  the  CSM
as given by  (\ref{densitaplummer})
when   $\eta=6$
produces the differential equation
\begin{equation}
\frac{d}{dt}r(t) = \frac {NDEP}{DNEP}
\quad ,
\label{eqndiffplummer}
\end{equation}
where
\begin{eqnarray}
NDEP =
 ( 9\,\sqrt {3}\arctan  ( 1/3\,{\frac {{\it r0}\,\sqrt {3}}{b
}}  ) {b}^{4}+6\,\sqrt {3}\arctan  ( 1/3\,{\frac {{\it r0}\,
\sqrt {3}}{b}}  ) {b}^{2}{{\it r0}}^{2} +
\nonumber \\
+\sqrt {3}\arctan  ( 1
/3\,{\frac {{\it r0}\,\sqrt {3}}{b}}  ) {{\it r0}}^{4}-9\,{b}^{3}
{\it r0}+3\,b{{\it r0}}^{3}  ) {\it v0}\,  ( 3\,{b}^{2}+
  ( r  ( t  )   ) ^{2}  ) ^{2}
\quad ,
\nonumber
\end{eqnarray}
and
\begin{eqnarray}
NDEP =
 ( 3\,{b}^{2}+{{\it r0}}^{2}  ) ^{2}  ( 9\,\sqrt {3}
\arctan  ( 1/3\,{\frac {r  ( t  ) \sqrt {3}}{b}}  )
{b}^{4}+6\,\sqrt {3}\arctan  ( 1/3\,{\frac {r  ( t  )
\sqrt {3}}{b}}  ) {b}^{2}  ( r  ( t  )   ) ^{2}
\nonumber \\
+\sqrt {3}\arctan  ( 1/3\,{\frac {r  ( t  ) \sqrt {3}}{b}
}  )   ( r  ( t  )   ) ^{4}-9\,{b}^{3}r  (
t  ) +3\,b  ( r  ( t  )   ) ^{3}  )
\quad .
\nonumber
\end{eqnarray}
There is no analytical solution to this differential equation,
but the  solution can be found numerically.

\subsection{Motion with Lane--Emden profile}

In the case of variable  density for  the  CSM
as given by  the profile (\ref{densita5}),
the differential equation
which models momentum conservation is
\begin{equation}
4\,{\frac {{b}^{3} ( r ( t )  ) ^{3}\rho_{{c}}
\pi \,\sqrt {3}{\frac {d}{dt}}r ( t ) }{ ( 3\,{b}^{2}+
 ( r ( t )  ) ^{2} ) ^{3/2}}}-4\,{\frac {{
b}^{3}{r_{{0}}}^{3}\rho_{{c}}\pi \,\sqrt {3}v_{{0}}}{ ( 3\,{b}^{2
}+{r_{{0}}}^{2} ) ^{3/2}}}=0
\quad .
\label{eqndiff}
\end{equation}
The variables can be separated and the solution  is
\begin{equation}
r(t;r_0,v_0,t_0,b) =
\frac{N}{D}
\quad  ,
\label{radiusemdent}
\end{equation}
where
\begin{eqnarray}
N  =
\sqrt {2}{r_{{0}}}^{3/4}\bigl ({r_{{0}}}^{13/2}+2\,{r_{{0}}}^{11/2}
 ( t-t_{{0}} ) v_{{0}}
\nonumber \\
+{r_{{0}}}^{9/2} ( t-t_{{0}}
 ) ^{2}{v_{{0}}}^{2}+6\,{b}^{2}{r_{{0}}}^{9/2}
\nonumber \\
+18\,{b}^{2}{r_{{0
}}}^{7/2} ( t-t_{{0}} ) v_{{0}}+\sqrt {A}{r_{{0}}}^{4}+
\sqrt {A}{r_{{0}}}^{3} ( t-t_{{0}} ) v_{{0}}
\nonumber   \\
+9\,{b}^{4}{r_{
{0}}}^{5/2}+36\,{b}^{4}{r_{{0}}}^{3/2}v_{{0}} ( t-t_{{0}}
 )
\nonumber \\
+9\,\sqrt {A}{b}^{2}{r_{{0}}}^{2}+18\,\sqrt {A}{b}^{4} \bigr )^{1/2}
\nonumber
\end{eqnarray}
and
\begin{eqnarray}
D=2\, ( 3\,{b}^{2}+{r_{{0}}}^{2} ) ^{3/2}
\nonumber
\quad  ,
\end{eqnarray}
with
\begin{eqnarray}
A(t-t_0)=
{r_{{0}}}^{3} ( t-t_{{0}} ) ^{2}{v_{{0}}}^{2}+36\,{b}^{4}
 ( t-t_{{0}} ) v_{{0}}
\nonumber  \\
+18\,{b}^{2}{r_{{0}}}^{2} ( t-t_
{{0}} ) v_{{0}}
\nonumber \\
+2\,{r_{{0}}}^{4} ( t-t_{{0}} ) v_{{0}
}+9\,{b}^{4}r_{{0}}+6\,{b}^{2}{r_{{0}}}^{3}+{r_{{0}}}^{5}
\nonumber
\quad  .
\end{eqnarray}
This is the {\it first} solution and  has an analytical form.
The  analytical  solution  for the velocity
can be found from  the first  derivative of the
analytical solution as represented by
equation~(\ref{radiusemdent}),
\begin{equation}
v(t;r_0,v_0,t_0,b) = \frac {d}{dt}r(t;r_0,v_0,t_0,b)
\quad  .
\label{velocityemden}
\end {equation}
The  previous  differential equation (\ref{eqndiff})
can be organized as
\begin{equation}
\frac {d}{dt} r ( t )  = f (r;r_0,v_0,t_0,b)
\quad ,
\label{eqnf}
\end{equation}
and we seek  a power series solution of the form
\begin{equation}
r(t) = a_0 +a_1  (t-t_0) +a_2  (t-t_0)^2+a_3  (t-t_0)^3 + \dots
\quad ,
\label{rtseries}
\end{equation}
see  \cite{Tenenbaum1963,Ince2012}.
The Taylor expansion of equation~(\ref{eqnf}) gives
\begin{eqnarray}
 f (r;r_0,v_0,t_0,b)=
\nonumber \\
b_0 +b_1  (t-t_0) +b_2  (t-t_0)^2+b_3  (t-t_0)^3 + \dots
\quad ,
\nonumber
\end{eqnarray}
where the values of $b_n$ are
\begin{eqnarray}
b_0=& f (r_0;r_0,v_0,t_0,b)                                    \nonumber \\
b_1=& \frac {\partial} {\partial t} f (r_0;r_0,v_0,t_0,b)      \nonumber \\
b_2=& \frac{1}{2!}\frac {\partial^2} {\partial t^2} f (r_0;r_0,v_0,t_0,b)  \\
b_3=& \frac{1}{3!}\frac {\partial^3} {\partial t^3} f (r_0;r_0,v_0,t_0,b)  \nonumber \\
\ldots &\dots \dots  \nonumber
\end{eqnarray}
The relation between the coefficients $a_n$ and $b_n$ is
\begin{eqnarray}
a_1=& b_0   \nonumber \\
a_2=& \frac{b_1}{2} \nonumber  \\
a_3=& \frac{b_2}{3}  \nonumber \\
\ldots &\dots \dots  \nonumber
\end{eqnarray}
The higher-order  derivatives plus the initial conditions give
\begin{eqnarray}
a_0=& r_0    \nonumber \\
a_1=& v_0   \nonumber \\
a_2=& -\frac{ 9\,{{\it v_0}}^{2}{b}^{2}} { 2\, ( 3\,{b}^{2}+{{\it r_0}}^{2} ) {\it r_0}}\nonumber  \\
a_3=&\frac{ 9\,{{\it v_0}}^{3}{b}^{2} ( 7\,{b}^{2}+{{\it r_0}}^{2} )  }{2\,{{\it r_0}}^{2} ( 3\,{b}^{2}+{{\it r_0}}^{2} ) ^{2} }  \\
\ldots &\dots \dots  \nonumber
\end{eqnarray}
These  are  the coefficient of the {\it second} solution,
which is a power series.

A {\it third } solution can be represented
by a difference equation
which has the following type of recurrence relation
\begin{eqnarray}
r_{n+1} =&  r_n + v_n \Delta t    \nonumber  \\
v_{n+1} =& \frac {{r_{{n}}}^{3}v_{{n}} ( 3\,{b}^{2}+{r_{{n+1}}}^{2} ) ^{3/2}} {( 3\,{b}^{2}+{r_{{n}}}^{2} ) ^{3/2}{r_{{n+1}}}^{3}}
\quad  ,
\label{recursive}
\end{eqnarray}
where  $r_n$, $v_n$, and $\Delta t$ are the temporary
radius,
the velocity, and the interval of time.

The physical units have not yet been specified:
pc for length  and  yr for time
are the units most commonly used by astronomers.
With these units, the initial velocity $v_{{0}}$
is  expressed in $\mathrm{pc \,yr^{-1}}$,
1 yr = 365.25 days,
and should be converted
into   km s$^{-1}$; this means
that   $v_{{0}} =1.02\times10^{-6} v_{{1}}$
where  $v_{{1}}$ is the initial
velocity expressed in
km s$^{-1}$.
In these units, the speed of light is
$c=0.306$  \ pc \ yr$^{-1}$.

\subsection{Motion with the Lane--Emden profile,
Pad\'e approximation}

Assuming a Lane--Emden profile, $n=5$,     as given by 
equation~(\ref {profile_lane}),
the velocity is
\begin{equation}
\frac{dr}{dt} = \frac{NL}{DL}
\quad ,
\label{vel_lane}
\end{equation}
where
\begin{eqnarray}
NL = {r_{{0}}}^{3}v_{{0}} \left( 3\,{b}^{2}+{r}^{2} \right) ^{\frac{3}{2}} \left( 3
\,{b}^{2}+{r_{{0}}}^{2} \right) ^{\frac{3}{2}}
\end{eqnarray}
and 
\begin{eqnarray}
DL = -3\, \left( 3\,{b}^{2}+{r}^{2} \right) ^{\frac{3}{2}}\sqrt {3}{r_{{0}}}^{3}{b}
^{3}+3\, \left( 3\,{b}^{2}+{r_{{0}}}^{2} \right) ^{\frac{3}{2}}\sqrt {3}{b}^{3
}{r}^{3}
\nonumber \\
+ \left( 3\,{b}^{2}+{r}^{2} \right) ^{\frac{3}{2}} \left( 3\,{b}^{2}+{
r_{{0}}}^{2} \right) ^{\frac{3}{2}}{r_{{0}}}^{3}
\, . 
\end{eqnarray}
The connected  nonlinear equation  is
\begin{eqnarray}
\frac {1}
{
{r_{{0}}}^{3}v_{{0}} \left( 3\,{b}^{2}+{r_{{0}}}^{2} \right) ^{\frac{3}{2}}
\sqrt {3\,{b}^{2}+{r}^{2}}
}
 \times 
\nonumber \\
\bigg (
54\,   ( {b}^{2}+\frac{1}{3}\,{r_{{0}}}^{2}   )    ( \frac{1}{18}\,{r_{{0}}}
^{3}   ( r-r_{{0}}   ) \sqrt {3\,{b}^{2}+{r}^{2}}
\nonumber \\
+{b}^{3}\sqrt 
{3}   ( {b}^{2}+\frac{1}{6}\,{r}^{2}   )    ) \sqrt {3\,{b}^{2}+{r_
{{0}}}^{2}}-54\,\sqrt {3\,{b}^{2}+{r}^{2}}\sqrt {3}{b}^{3}   ( {b}^
{4}
\nonumber \\ 
+\frac{1}{2}\,{b}^{2}{r_{{0}}}^{2}+\frac{1}{18}\,r{r_{{0}}}^{3}   )
\bigg ) =t-t_0
\nonumber  
\quad .
\end{eqnarray}
The Pad\'e rational polynomial 
approximation of degree 2 in the numerator
and degree 1 in the denominator for the left-hand side of the above equation gives
\begin{eqnarray}
\frac
{
NP
}
{
2\, \left( 3\,{b}^{2}+{r_{{0}}}^{2} \right) ^{\frac{3}{2}}v_{{0}} \left( 2\,r{
b}^{2}-5\,{b}^{2}r_{{0}}-r{r_{{0}}}^{2} \right) 
}
=t-t_0
\, ,
\end{eqnarray}
where 
\begin{eqnarray}
PN =
-27\,   ( r-r_{{0}}   )  
  \Big ( \big  ( -\frac{4}{9}\,   ( r{b}^{2}-\frac{5}{2}\,{b}
^{2}r_{{0}}-\frac{1}{2}\,r{r_{{0}}}^{2} \big  ) \times
\nonumber \\
  \big ( {b}^{2}+\frac{1}{3}\,{r_{{0}}}
^{2}  \big ) \sqrt {3\,{b}^{2}+{r_{{0}}}^{2}}+{b}^{5}\sqrt {3}  \big ( 
r-r_{{0}} \big  )   \Big  ) 
\, .
\end{eqnarray}
The  Pad\'e  approximant  for the radius is 
\begin{eqnarray}
r_{2,1}=\frac{NR}{DR}
\label{rmotionlaneemden}   
\end{eqnarray}
where 
\begin{eqnarray}
NR= -18\,   ( {b}^{2}+\frac{1}{3}\,{r_{{0}}}^{2}    ) ^{2}{b}^{2} 
  ( -\frac{1}{2}\,{r_{{0}}}^{3}-\frac{1}{2}\,v_{{0}}   ( t-t_{{0}}    ) {r_{{0}}}^{2}
\nonumber \\
+ \frac{7}{2} \,{b}^{2}r_{{0}}+{b}^{2}v_{{0}}   ( t-t_{{0}}    )     ) 
\sqrt {3\,{b}^{2}+{r_{{0}}}^{2}}+   ( 81\,{b}^{9}r_{{0}}+27\,{b}^{7
}{r_{{0}}}^{3}    ) \sqrt {3}
\nonumber \\
+\sqrt {972}
\Bigg ( {   ( {b}^{2}+\frac{1}{3}
\,{r_{{0}}}^{2}    ) ^{4}{b}^{4}   ( \frac{9}{2}\,\sqrt {3}r_{{0}}{b}^{5
}v_{{0}}   ( t-t_{{0}}    ) \sqrt {3\,{b}^{2}
+{r_{{0}}}^{2}}
}
\nonumber \\
{
+
 \bigg  ( -\frac{1}{2}\,{r_{{0}}}^{3}-\frac{1}{2}\,v_{{0}}   ( t-t_{{0}}    ) {r_{
{0}}}^{2}-\frac{3}{2}\,{b}^{2}r_{{0}}
}
\nonumber \\
{
+{b}^{2}v_{{0}}   ( t-t_{{0}}    ) 
    ) ^{2}   ( {b}^{2}+\frac{1}{3}\,{r_{{0}}}^{2}    )  \bigg   ) } \Bigg )^{\frac{1}{2}}
\, ,
\end{eqnarray}
and  
\begin{eqnarray}
DR=
{b}^{2}   ( 3\,{b}^{2}+{r_{{0}}}^{2}   ) \bigg   ( 27\,{b}^{5}
\sqrt {3}-12\,{b}^{4}\sqrt {3\,{b}^{2}+{r_{{0}}}^{2}}
\nonumber \\
+2\,{b}^{2}{r_{{0
}}}^{2}\sqrt {3\,{b}^{2}+{r_{{0}}}^{2}}+2\,{r_{{0}}}^{4}\sqrt {3\,{b}^
{2}+{r_{{0}}}^{2}} \bigg  ) 
\quad ,
\end{eqnarray}
and the velocity is
\begin{equation}
v_{2,1}=\frac{dr_{2,1}}{dt} =\frac{NVL}{DVL}
\quad ,
\label{vmotionlaneemden}
\end{equation}
where 
\begin{eqnarray} 
NVL =
-18\,\sqrt {3}   ( 3\,{b}^{2}+{r_{{0}}}^{2}   ) v_{{0}} 
  \bigg  (  
  \bigg  ( -243\,   ( {b}^{2}+\frac{1}{3}\,{r_{{0}}}^{2}   ) ^{2}{b}^{7}r_
{{0}}\sqrt {3}
\nonumber \\
+\sqrt {972}
{ \Bigg \{  ( {b}^{2}+\frac{1}{3}\,{r_{{0}}}^{2}
   ) ^{4}{b}^{4}   ( 9/2\,\sqrt {3}r_{{0}}{b}^{5}v_{{0}}
   ( t-t_{{0}}   ) \sqrt {3\,{b}^{2}+{r_{{0}}}^{2}}
}
\nonumber \\
{
+   ( {b}
^{2}+\frac{1}{3}\,{r_{{0}}}^{2}   ) \bigg   ( -\frac{1}{2}\,{r_{{0}}}^{3}-\frac{1}{2}\,v_{{0
}}   ( t-t_{{0}}   ) {r_{{0}}}^{2}-3/2\,{b}^{2}r_{{0}}
}
\nonumber  \\
{
+{b}^{2}v
_{{0}}   ( t-t_{{0}}   )    ) ^{2} \bigg  ) } 
\Bigg \}^{\frac{1}{2}}
  ( 2\,{b}^
{2}-{r_{{0}}}^{2}   ) \Bigg   ) \sqrt {3\,{b}^{2}+{r_{{0}}}^{2}}
\nonumber  \\
-
108\,   ( {b}^{2}+\frac{1}{3}\,{r_{{0}}}^{2}   ) ^{3}{b}^{2}   ( -1/
2\,{r_{{0}}}^{3}-\frac{1}{2}\,v_{{0}}   ( t-t_{{0}}   ) {r_{{0}}}^{2}-3
/2\,{b}^{2}r_{{0}}
\nonumber  \\
+{b}^{2}v_{{0}}   ( t-t_{{0}}   )    ) 
   ( {b}^{2}-\frac{1}{2}\,{r_{{0}}}^{2} ) 
\quad ,
\end{eqnarray}
and
\begin{eqnarray} 
DVL =
18\,\sqrt {972}\sqrt {3}
\Bigg 
\{ 
  ( {b}^{2}+\frac{1}{3}\,{r_{{0}}}^{2}
   ) ^{4}{b}^{4}   ( 9/2\,\sqrt {3}r_{{0}}{b}^{5}v_{{0}}
\nonumber \\
   ( t-t_{{0}}   ) \sqrt {3\,{b}^{2}+{r_{{0}}}^{2}}+   ( {b}
^{2}+\frac{1}{3}\,{r_{{0}}}^{2}   )    ( -\frac{1}{2}\,{r_{{0}}}^{3}-\frac{1}{2}\,v_{{0
}}   ( t-t_{{0}}   ) {r_{{0}}}^{2}
\nonumber  \\
-3/2\,{b}^{2}r_{{0}}+{b}^{2}v
_{{0}}   ( t-t_{{0}}   )    ) ^{2}   ) 
\Bigg 
\}
^{\frac{1}{2}} 
  ( 
   ( -12\,{b}^{4}+2\,{b}^{2}{r_{{0}}}^{2}+2\,{r_{{0}}}^{4}   ) 
\sqrt {3\,{b}^{2}+{r_{{0}}}^{2}}
\nonumber \\
+27\,{b}^{5}\sqrt {3}   ) 
\quad .
\end{eqnarray}

\subsection{Motion with a power law profile}

The differential equation which models momentum conservation
in the presence of a power law behavior of the density,
as given by (\ref{piecewise}),
is
\begin{eqnarray}
( -4\,{\frac { ( r ( t )  ) ^{3}\rho_{{c}}
\pi }{d-3} ( {\frac {r_{{0}}}{r ( t ) }} ) ^{d}}
+4\,{\frac {\rho_{{c}}\pi \,{r_{{0}}}^{3}}{d-3}}+4/3\,\rho_{{c}}\pi \,
{r_{{0}}}^{3} ) {\frac {\rm d}{{\rm d}t}}r ( t )
\nonumber \\
-4/3 \,\rho_{{c}}\pi \,{r_{{0}}}^{3}{\it v_0}=0
\quad .
\label{eqndiffpowerlaw}
\end{eqnarray}

A {\it first} solution can be found numerically,
see \cite{Zaninetti2011a} for more details.
A {\it second} solution is a  truncated series
about the ordinary point $t=t_0$
which to fourth order has
coefficients
\begin{eqnarray}
a_0=& r_0    \nonumber \\
a_1=& v_0     \nonumber \\
a_2=&  \frac{-3\,{{\it v0}}^{2}} {2\,{\it r_0}} \nonumber \\
a_3=&   \frac{ ( d+7 ) {{\it v_0}}^{3}  } {2\,{{\it r_0}}^{2} } \quad .
\label{coefficientspower}
\end{eqnarray}
A {\it third} approximate solution  can be
found
assuming that
$3 r_0^d r^{4-d}$
$\gg$
$-(4 r_0^3 d-r_0^3 d^2)r$
\begin{eqnarray}
 r(t) =
 ( {r_{{0}}}^{4-d}-\frac{1}{3}d{r_{{0}}}^{4-d}
( 4-d ) \nonumber \\
 + \frac{1}{3}
 ( 4-d ) v_{{0}}{r_{{0}}}^{3-d} ( 3-d )
 ( t-t_{{0}} )  ) ^{\frac{1}{4-d}}
\quad .
\nonumber
\label{asymptotic}
\end{eqnarray}
This is an important approximate result because,
given  the astronomical relation
$r(t)\propto t^{\alpha}$,
we have  $d=4 -\frac{1}{\alpha}$.

\section{Conservation of the relativistic momentum }

\label{secrelativistic}
The  thin layer approximation assumes
that all the swept  mass
during the travel from the initial time, $t_0$,
to the time $t$, resides
in a thin shell of radius $r(t)$ with velocity $v(t)$.
The relativistic  conservation of momentum,
see \cite{French1968,Zhang1997,Guery2010},
is formulated as
\begin{equation}
M(r_0) \gamma_0 \beta_0 = M(r) \gamma \beta
\quad ,
\nonumber
\end{equation}
where
\begin{equation}
\gamma_0 = \frac{1} {
\sqrt{1-\beta_0^2}
}
\quad ; \qquad
\gamma = \frac{1} {
\sqrt{1-\beta^2}
}
\quad ,
\nonumber
\end {equation}
and
\begin{equation}
\beta_0 =\frac{v_0}{c}
\quad ; \qquad
\beta =\frac{v}{c}.
\nonumber
\end{equation}

\subsection{Lane--Emden case ($n=5$)}

On assuming   a Lane--Emden dependence ($n=5$),
the total  mass $M(r;b)$ contained between 0 and  $r$
is given by equation ~(\ref{integralemassa}).
The equation of the relativistic conservation of momentum
is easily solved for  $\beta$ as a function
of the radius:
\begin{equation}
\beta = \frac{
\sqrt {A ( 3\,{b}^{2}+{r}^{2} ) } ( 3\,{b}^{2}+{r}^{2}
 ) {r_{{0}}}^{3}\beta_{{0}}
 }
 {A}
\quad ,
\nonumber
\label{eqnbeta}
\end{equation}
with
\begin{eqnarray}
A = -27\,{b}^{6}{r}^{6}{\beta_{{0}}}^{2}+27\,{b}^{6}{\beta_{{0}}}^{2}{r_{{0
}}}^{6}-27\,{b}^{4}{r}^{6}{\beta_{{0}}}^{2}{r_{{0}}}^{2}+27\,{b}^{4}{r
}^{2}{\beta_{{0}}}^{2}{r_{{0}}}^{6}
\nonumber \\
-9\,{b}^{2}{r}^{6}{\beta_{{0}}}^{2}
{r_{{0}}}^{4}+9\,{b}^{2}{r}^{4}{\beta_{{0}}}^{2}{r_{{0}}}^{6}+27\,{b}^
{6}{r}^{6}+27\,{b}^{4}{r}^{6}{r_{{0}}}^{2}+9\,{b}^{2}{r}^{6}{r_{{0}}}^
{4}+{r}^{6}{r_{{0}}}^{6}
\nonumber
\quad .
\end{eqnarray}
Inserting
\begin{equation}
\beta = \frac{1}{c}{\frac {d}{dt}}r( t ),
\nonumber
\end{equation}
the relativistic conservation of momentum
can be written as the differential
equation
\begin{equation}
\frac{
4\,{b}^{3} ( r ( t )  ) ^{3}\rho\,\pi \,\sqrt {3
}{\frac {d}{dt}}r ( t )
}
{
( 3\,{b}^{2}+ ( r ( t )  ) ^{2} ) ^{
3/2}c\sqrt {-{\frac { ( {\frac {d}{dt}}r ( t )
 ) ^{2}}{{c}^{2}}}+1}
}
= \frac {
4\,{b}^{3}{r_{{0}}}^{3}\rho\,\pi \,\sqrt {3}\beta_{{0}}
}
{
( 3\,{b}^{2}+{r_{{0}}}^{2} ) ^{3/2}\sqrt {-{\beta_{{0}}}^{
2}+1}
}
\quad .
\label{eqndiffrel}
\end{equation}
This first order differential equation
can be solved by separating the variables:
\begin{equation}
\int_{r_0}^r
\frac{A}{
\sqrt {A ( 3\,{b}^{2}+{r}^{2} ) } ( 3\,{b}^{2}+{r}^{2}
 ) {r_{{0}}}^{3}\beta_{{0}}
 }
\,dr
= c (t-t_0)
\quad .
\label{nlequation}
\end{equation}
The previous integral does not have an analytical solution
and we treat the previous result as a nonlinear
equation to be
solved numerically.
The differential equation has a truncated series solution
about the ordinary point $t=t_0$
which to fifth order is
\begin{equation}
r_s( t )=
\sum _{n=0}^{4}a_{{n}}{(t-t_0)}^{n}
\quad .
\label{rtseriesrel}
\end{equation}
The coefficients are
\begin{eqnarray}
a_0=& r_0    \nonumber \\
a_1=& c\beta_{{0}}   \nonumber \\
a_2=& \frac {9\,{b}^{2}{\beta_{{0}}}^{2}{c}^{2} ( {\beta_{{0}}}^{2}-1 )}{ 2\,r_{{0}} ( 3\,{b}^{2}+{r_{{0}}}^{2} ) }              \\
a_3=&\frac{9\,{c}^{3} ( \beta_{{0}}-1 )  ( \beta_{{0}}+1 )
 ( 12\,{b}^{2}{\beta_{{0}}}^{2}-7\,{b}^{2}-{r_{{0}}}^{2} )
{\beta_{{0}}}^{3}{b}^{2}} {2\, ( 3\,{b}^{2}+{r_{{0}}}^{2} ) ^{2}{r_{{0}}}^{2}} \nonumber \\
a_4=&\frac{9\,{b}^{2} ( \beta_{{0}}-1 )  ( \beta_{{0}}+1 )
B{\beta_{{0}}}^{4}{c}^{4}} {8\,{r_{{0}}}^{3} ( 3\,{b}^{2}+{r_{{0}}}^{2} ) ^{3}} \nonumber  \\
where ~B=& 756\,{b}^{4}{\beta_{{0}}}^{4}-927\,{b}^{4}{\beta_{{0}}}^{2}-117\,{b}^{
2}{\beta_{{0}}}^{2}{r_{{0}}}^{2}+231\,{b}^{4}+69\,{b}^{2}{r_{{0}}}^{2}
+4\,{r_{{0}}}^{4}  \nonumber
\end{eqnarray}
The velocity approximated to the fifth order is
\begin{equation}
v_s( t ) =
\sum _{n=1}^{4}{\frac {a_{{n}}{(t-t_0)}^{n}n}{(t-t_0)}}
\quad .
\label{vseries}
\end{equation}
The presence of an analytical  expression for $\beta$
as given  by equation~(\ref{eqnbeta})  allows the
recursive solution
\begin{eqnarray}
r_{n+1} =   &  r_n + c \beta_n \Delta t    \nonumber  \\
\beta_{n+1} =&
\frac
{
\sqrt {A ( 3\,{b}^{2}+{r_{{n+1}}}^{2} ) } ( 3\,{b}^{2}
+{r_{{n+1}}}^{2} ) {r_{{n}}}^{3}\beta_{{n}}
}
{
A
}    \\
with ~A=&
27\,{b}^{6}{\beta_{{n}}}^{2}{r_{{n}}}^{6}-27\,{b}^{6}{\beta_{{n}}}^{2}
{r_{{n+1}}}^{6}+27\,{b}^{4}{\beta_{{n}}}^{2}{r_{{n+1}}}^{2}{r_{{n}}}^{
6}
\nonumber    \\
-27\,{b}^{4}{\beta_{{n}}}^{2}{r_{{n+1}}}^{6}{r_{{n}}}^{2}
& +9\,{b}^{2}
{\beta_{{n}}}^{2}{r_{{n+1}}}^{4}{r_{{n}}}^{6}-9\,{b}^{2}{\beta_{{n}}}^
{2}{r_{{n+1}}}^{6}{r_{{n}}}^{4}+27\,{b}^{6}{r_{{n+1}}}^{6}
\nonumber  \\
+27\,{b}^{4}
{r_{{n+1}}}^{6}{r_{{n}}}^{2}
& +9\,{b}^{2}{r_{{n+1}}}^{6}{r_{{n}}}^{4}+{r
_{{n+1}}}^{6}{r_{{n}}}^{6}
\nonumber
\quad  ,
\label{recursiverel}
\end{eqnarray}
where  $r_n$, $\beta_n$ and $\Delta t$
are the temporary  radius,
the relativistic $\beta$
factor, and the interval of time, respectively.
Up to now we have taken the time interval $t-t_0$ to be that as seen
by an observer on earth.
For an   observer which moves on the expanding shell,
the proper time $\tau^*$ is
\begin{equation}
\tau^* = \int_{t_0}^{t} \frac {dt}{\gamma} =  \int_{t_0}^{t}
\sqrt{1-\beta^2} dt
\quad ,
\nonumber
\end{equation}
see  \cite{Larmor1897,Lorentz1904,Einstein1905,Macrossan1986}.
In  the series solution framework,
$\beta=\frac{v_s}{c}$,
and   $v_s$ is given by equation~(\ref{vseries}).

\subsection{Plummer case ($\eta=6$)} 

The relativistic  conservation of momentum
for a Plummer profile with $\eta=6$ is 
\begin{equation}
\frac{AN}{AD} = \frac{BN}{BD}
\quad , 
\label{eqndiffrelplummer}
\end{equation}
\begin{eqnarray}
AN=
3\,  \,\pi\,{b}^{3}  \Bigl  ( 9\,\arctan   ( \frac{1}{3}\,{\frac {r   ( t
    ) \sqrt {3}}{b}}    ) \sqrt {3}{b}^{4}+6\,\arctan   ( \frac{1}{3}
\,{\frac {r   ( t    ) \sqrt {3}}{b}}    ) \sqrt {3}{b}^{2}
   ( r   ( t    )     ) ^{2}
\nonumber  \\
+\arctan   ( \frac{1}{3}\,{\frac {r
   ( t    ) \sqrt {3}}{b}}    ) \sqrt {3}   ( r   ( t
    )     ) ^{4}-9\,{b}^{3}r   ( t    ) +3\,b   ( r
   ( t    )     ) ^{3}   \Bigr ) {\frac {\rm d}{{\rm d}t}}r
   ( t    ) 
\end {eqnarray}
\begin{equation}
AD=
2\, \left( 3\,{b}^{2}+ \left( r \left( t \right)  \right) ^{2}
 \right) ^{2}c\sqrt {1-{\frac { \left( {\frac {\rm d}{{\rm d}t}}r
 \left( t \right)  \right) ^{2}}{{c}^{2}}}}
\quad  , 
\end{equation}

\begin{eqnarray}
BN=
3\,  \,\pi\,{b}^{3}  \Bigl  ( 9\,\arctan    ( \frac{1}{3}\,{\frac {R_{{0}}
\sqrt {3}}{b}}    ) \sqrt {3}{b}^{4}+6\,\arctan    ( \frac{1}{3}\,{\frac 
{R_{{0}}\sqrt {3}}{b}}    ) \sqrt {3}{b}^{2}{R_{{0}}}^{2}
\nonumber  \\
+\arctan
    ( \frac{1}{3}\,{\frac {R_{{0}}\sqrt {3}}{b}}    ) \sqrt {3}{R_{{0}}}^
{4}
-9\,{b}^{3}R_{{0}}+3\,b{R_{{0}}}^{3}  \Bigr  ) \beta_{{0}}
\quad  , 
\end{eqnarray}
\begin{equation}
BD=
2\, \left( 3\,{b}^{2}+{R_{{0}}}^{2} \right) ^{2}\sqrt {1-{\beta_{{0}}}^
{2}}
\quad .
\end{equation}
The above differential equation has a truncated series solution
about the ordinary point $t=t_0$
which to fourth order is
\begin{equation}
r_s( t )=
\sum _{n=0}^{3}a_{{n}}{(t-t_0)}^{n}
\quad .
\label{rtseriesrelplummer6}
\end{equation}
The coefficients are
\begin{eqnarray}
a_0=& r_0    \nonumber \\
a_1=& 36\,{c}^{2} \left( {\beta0}^{2}-1 \right) {b}^{3}{{\it r0}}^{2}{\beta0
}^{2} \nonumber \\
a_2= \frac{A2N}{A2D}
\quad  ,
\end{eqnarray}
where  
\begin{equation}
A2N = 36\,{c}^{2} \left( {\beta0}^{2}-1 \right) {b}^{3}{{\it r0}}^{2}{\beta0
}^{2}
\quad  ,
\end{equation}
and  
\begin{eqnarray}
A2D 
=
27\,\sqrt {3}{b}^{6}\arctan \left( \frac{1}{3}\,{\frac {{\it r0}\,\sqrt {3}}{b
}} \right) +27\,\sqrt {3}{b}^{4}{{\it r0}}^{2}\arctan \left( \frac{1}{3}\,{
\frac {{\it r0}\,\sqrt {3}}{b}} \right) 
\nonumber \\
+9\,\sqrt {3}{b}^{2}{{\it r0}}
^{4}\arctan \left( \frac{1}{3}\,{\frac {{\it r0}\,\sqrt {3}}{b}} \right) +
\sqrt {3}{{\it r0}}^{6}\arctan \left( \frac{1}{3}\,{\frac {{\it r0}\,\sqrt {3}
}{b}} \right)
\nonumber \\
 -27\,{b}^{5}{\it r0}+3\,b{{\it r0}}^{5}
\quad .
\end{eqnarray}

\section{Classical  applications to \snr}

\label {applicationsclassic}
This section introduces:
the SN chosen for testing purposes,
the astrophysical environment connected with the selected SN,
two types of fit
commonly used to model the radius--time relation
in SN,
and the
application of the results obtained for
the Lane--Emden,
Plummer, and 
power law profiles.

\label{secobservations}

\subsection{The data}

\begin{figure*}
\begin{center}
\includegraphics[width=7cm ]{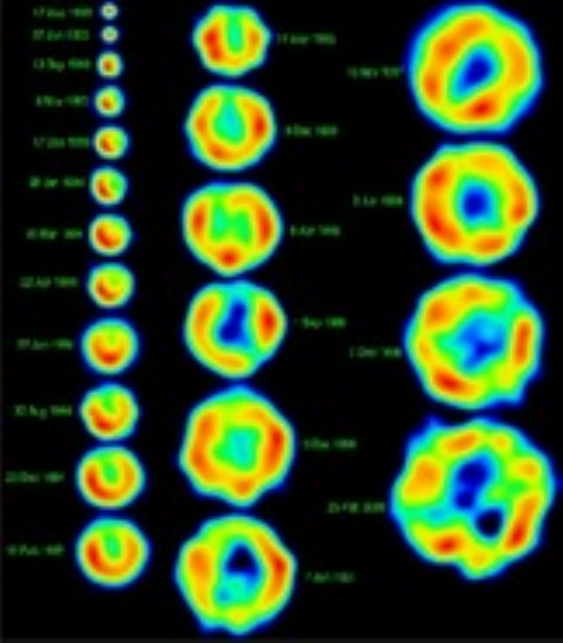}
\end {center}
\caption
{
Radio images of \snr as  function of time.
The color scale represents the brightness of the radio emission, 
with blue
being faintest and red brightest.
}
\label{SN1993J}
    \end{figure*}

The data of \snr,
radius in  pc and elapsed time  in years,
can be found in  Table 1 of  \cite{Marcaide2009}
and a multiple radio image is displayed in 
Figure \ref{SN1993J}.

The instantaneous velocity of expansion
can be deduced from the formula
\begin{equation}
v_i = \frac
{r_{i+1} - r_i}
{t_{i+1} - t_i}
\quad ,
\nonumber
\label{vdiscrete}
\end{equation}
where $r_i$ is the radius and $t_i$  is
the time
at the position $i$.
The uncertainty in the instantaneous   velocity
is found by implementing the error
propagation equation, see \cite{Bevington2003}.
A discussion of the
thickness of the radio shell
in \snr in the framework of
a reverse shock \cite{Chevalier1982a,Chevalier1982b}
can be found in \cite{Bartel2007}.
The thickness of the radio shell can also be explained
in the framework of the image theory, see
Section 6.3 in \cite{Zaninetti2011a}.

\subsection{Astrophysical Scenario}

The progenitor of \snr  was a K-supergiant star, see \cite{Aldering1994}
 and probably formed a binary system with a B-supergiant companion
star, see \cite{Maund2004}.
These massive stars have strong
stellar winds, and blow huge bubbles (of $\approx$  20 to 40 pc in size) in their
lives.
>From an analytic approximation, \cite{Weaver1977}
obtained a formula for the radius of the bubble, their eq. (21),
see too their Fig. 3.
The inner region of the bubble has very low density, and the border of the bubble
is the wall of a relatively dense shell which is in contact with the ISM.
The circumstellar envelope of the Pre \snr with which the
SN shock front is interacting is a small structure within the big bubble
created by the strong
stellar winds of the SN progenitor
(and probably of its binary companion) during
its life. Therefore this envelope of the Pre \snr
would be the product of a
recent event of stellar mass ejection suffered by the Pre \snr.
That is to
say that the SN shock wave interacts with a CSM created by
Pre supernova mass loss. In this respect,
\cite{Schmidt1993}
gave evidence that significant mass loss
 had taken place before the explosion, see also \cite{Smith2008a}.
 In the scenario in which the Pre \snr
had formed an interacting binary system,
this can be interpreted in terms of a
process of mass transfer.
It is possible that this type of supernova originates
in interacting binary systems.

\subsection{Two types of fit}

The quality of the fits is measured by the
merit function
$\chi^2$
\begin{equation}
\chi^2  =
\sum_j \frac {(r_{th} -r_{obs})^2}
             {\sigma_{obs}^2}
\quad ,
\nonumber
\label{chisquare}
\end{equation}
where  $r_{th}$, $r_{obs}$ and $\sigma_{obs}$
are the theoretical radius, the observed radius, and
the observed uncertainty, respectively.
A {\em first} fit
can be done by assuming  a  power law
dependence  of the type
\begin{equation}
r(t) = r_p t^{\alpha_p}
\nonumber
\label{rpower}
\quad ,
\end{equation}
where the  two parameters $r_p$ and  $\alpha_p$
as well their  uncertainties
can be found
using the recipes  suggested in
\cite{Zaninetti2011a}.
A {\em second}  fit   can be done by assuming
a piecewise function as  in
Fig.~4 of \cite{Marcaide2009}
\begin{equation}
 r(t)   = \{ \begin{array}{ll}
             r_{br}(\frac{t}{t_{br}})^{\alpha_1} &
              \mbox {if $t \leq t_{br} $ } \nonumber \\
             r_{br}(\frac{t}{t_{br}})^{\alpha_2} &
               \mbox {if $t >; t_{br} $. } \nonumber
            \end{array}
            .
\label{piecewisefit}
\end{equation}
This type of fit requires the determination
of four parameters: $t_{br}$ the break time,
$r_{br}$ the radius of expansion at
$t=t_{br}$,
and the exponents $\alpha_1$ and $\alpha_2$ of the two phases.
 The parameters of these two
 fits  as well the $\chi^2$  can  be found in
 Table~\ref{datafit}.
\begin{table}
\caption
{
Numerical values of the parameters
of the fits and
$\chi^2$;
$N$  represents the number of
free parameters.
}
 \label{datafit}
 \[
 \begin{array}{cccc}
 \hline
 \hline
 \noalign{\smallskip}
  N
& values  & \chi^2        \\
 \noalign{\smallskip}
 \hline
 \noalign{\smallskip}
  &\mathrm{power~law~as~a~fit}  &   &\\ \noalign{\smallskip}
  2   & \alpha_p = 0.82 \pm 0.0048  & 6364  \\ \noalign{\smallskip}
 ~    &r_p = (0.015 \pm 0.00011) ~{\mathrm{pc}}  &   \\
 \noalign{\smallskip}
 \hline
        & \mathrm{piecewise~fit}   &   &\\ \noalign{\smallskip}
 4   & \alpha_1 = 0.83 \pm 0.01  & 32

 \\
   & \alpha_2 = 0.78 \pm 0.0077; & ~
 \\

 \noalign{\smallskip}
 ~   & r_{br} = 0.05~{\mathrm{pc}};
 t_{br}=4.10~{yr}
&
 \\
 \noalign{\smallskip}
 \hline
  &\mathrm{Plummer~profile},~\eta=6  &   &\\ \noalign{\smallskip}
  2   & b=0.0045~{\mathrm{pc}} ; r_{0} = 0.008~{\mathrm{pc}};v_0=19500\frac{km}{s} &  265
\\ \noalign{\smallskip}
 \hline
  &\mathrm{Lane--Emden~profile}  &   &\\ \noalign{\smallskip}
  2   & b=0.00367~{\mathrm{pc}}; r_{0} = 0.008~{\mathrm{pc}};v_0=19500\frac{km}{s} &  471
\\
\noalign{\smallskip} \hline
  &\mathrm{Power~law~profile} &   &\\ \noalign{\smallskip}
  2   & d=2.93; r_{0} = 0.0022~{\mathrm{pc}}; &  276
\\
\noalign{\smallskip} ~   & t_{0}=0.249~{yr}; v_0=100000\frac{km}{s} & ~
\\
\noalign{\smallskip} \hline\hline
 \end{array}
 \]
 \end {table}

\subsection{The Lane--Emden  case}

\begin{figure*}
\begin{center}
\includegraphics[width=7cm ]{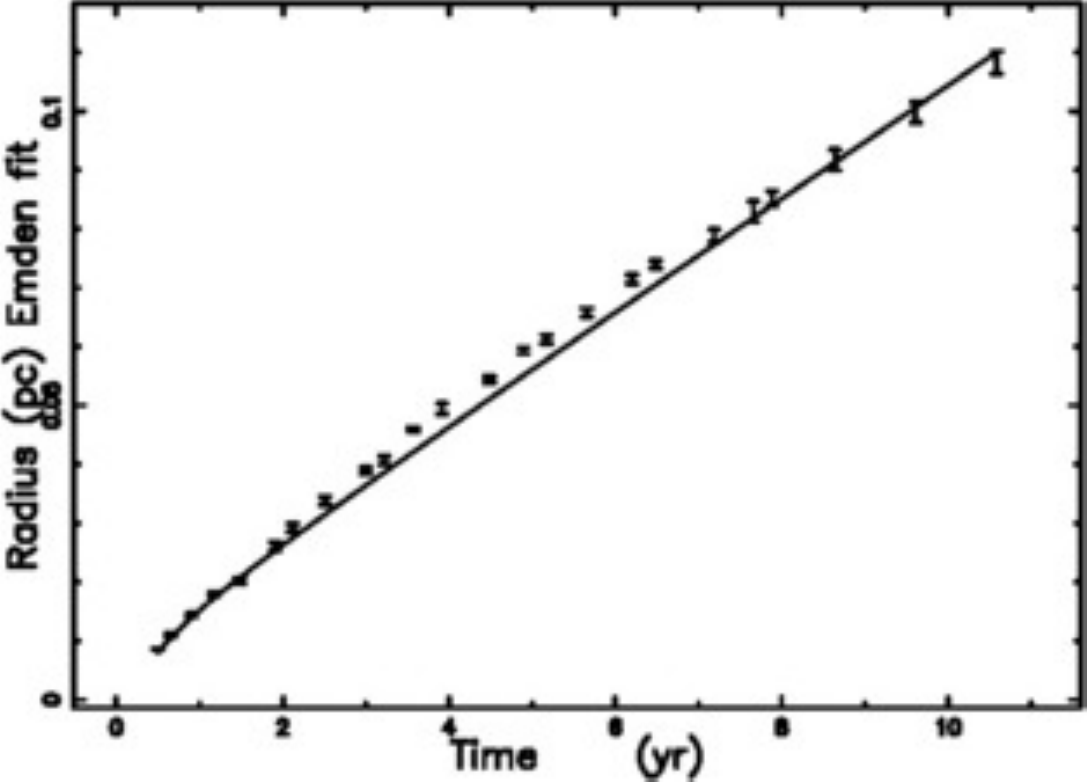}
\end {center}
\caption
{
Theoretical radius as given
by equation~(\ref{radiusemdent})
(full line),
with data as in Table~\ref{datafit}.
The  astronomical data of \snr
is represented with vertical error bars.
}
\label{1993pc_fit_emden}
    \end{figure*}

The radius  of \snr
which represents the  momentum conservation
in a Lane--Emden profile  of density
is  shown  in Figure~\ref{1993pc_fit_emden};
$r_0$ and $t_0$ are  fixed by the observations
and the two free  parameters are $b$ and $v_0$.

\begin{figure*}
\begin{center}
\includegraphics[width=7cm ]{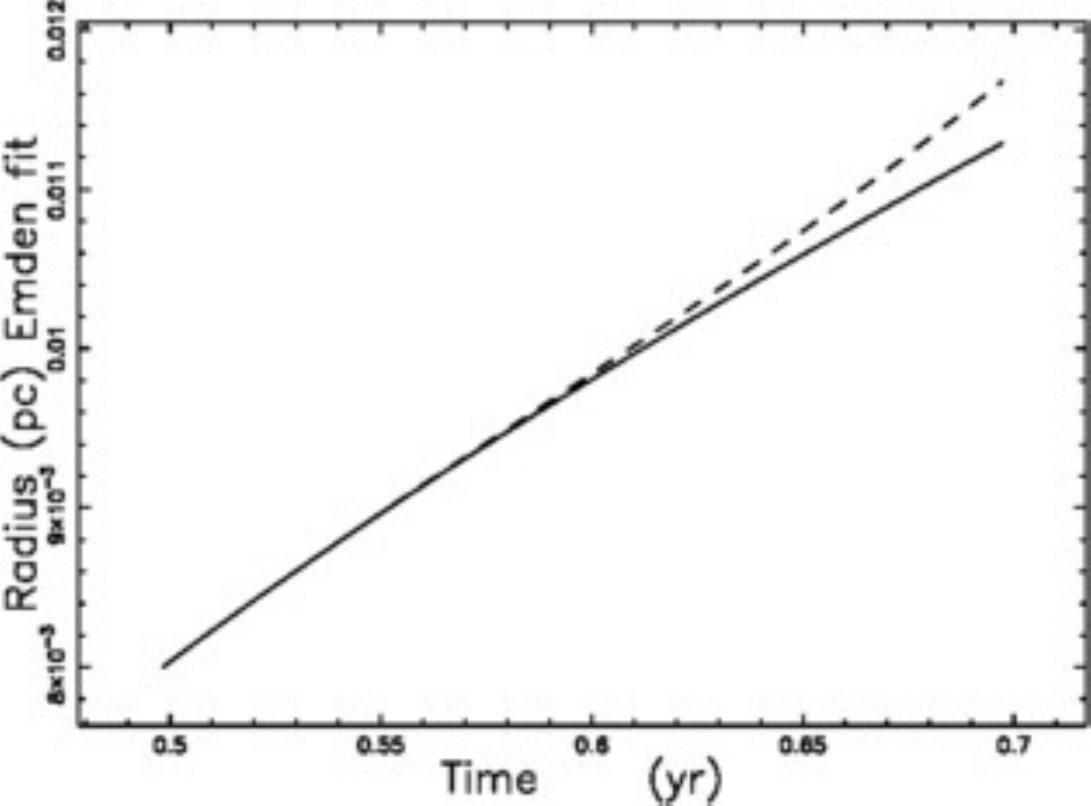}
\end {center}
\caption
{
Theoretical radius as given
by equation~(\ref{radiusemdent}) (full line)
and series solution as
given by equation~(\ref{rtseries}) (dashed line).
Data as in Table \ref{datafit}.
}
\label{1993pc_fit_series}
    \end{figure*}

Figure~\ref{1993pc_fit_series} compares
the theoretical solution and the series expansion
about the  ordinary point $t_0$.

\begin{figure*}
\begin{center}
\includegraphics[width=7cm ]{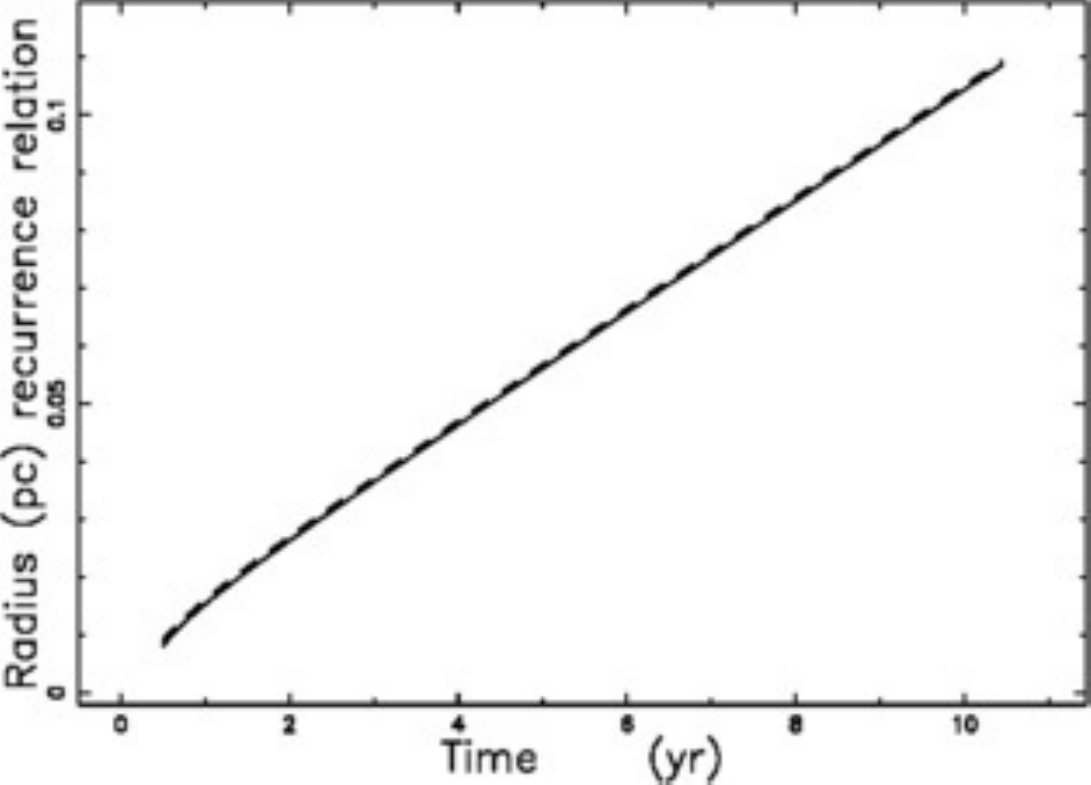}
\end {center}
\caption
{
Theoretical radius as given
by equation~(\ref{radiusemdent}) (full line)
and recursive solution as
given by equation~(\ref{recursive})
when $\Delta t=0.05 \mathrm{yr}$ (dashed line).
Data as in Table \ref{datafit}.
}
\label{1993pc_fit_recurs}
    \end{figure*}

The range of time in which  the series solution
approximates the analytical solution is limited.
Figure~\ref{1993pc_fit_recurs} compares
the theoretical solution and
the recursive solution
as represented by equation~\ref{recursive}.

\begin{figure*}
\begin{center}
\includegraphics[width=7cm ]{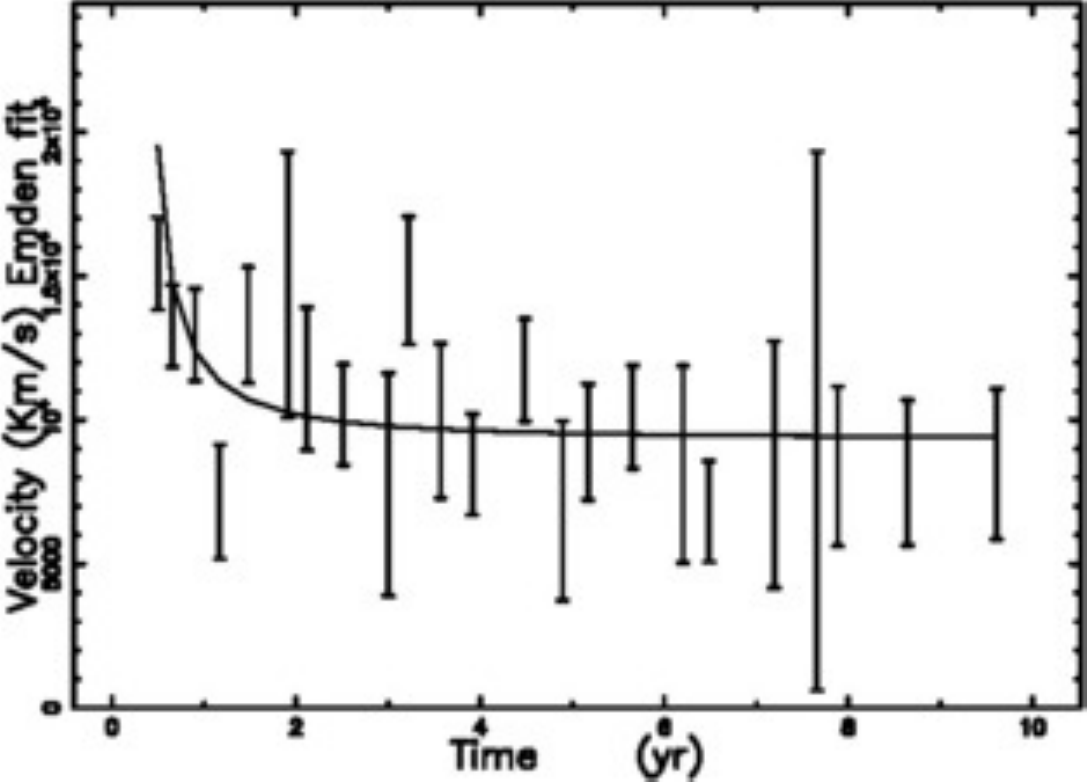}
\end {center}
\caption
{
Instantaneous velocity
of \snr with
uncertainty  and    theoretical  velocity
as given by  equation~(\ref{velocityemden}) (full line).
Data as in Table \ref{datafit}.
}
\label{1993pc_velocity_emden}
    \end{figure*}

The recursive solution
approximates the analytical solution
over the entire range of time
considered, and the error at $t=10$\ yr is $\approx$ 0.6$\%$ when
$\Delta t=0.05$ yr and $\approx$ 0.1$\%$ when
$\Delta t=0.0083$ yr.
The time evolution of the velocity is shown
in Figure~\ref{1993pc_velocity_emden}.

\subsection{Plummer and power law cases}

\begin{figure*}
\begin{center}
\includegraphics[width=7cm ]{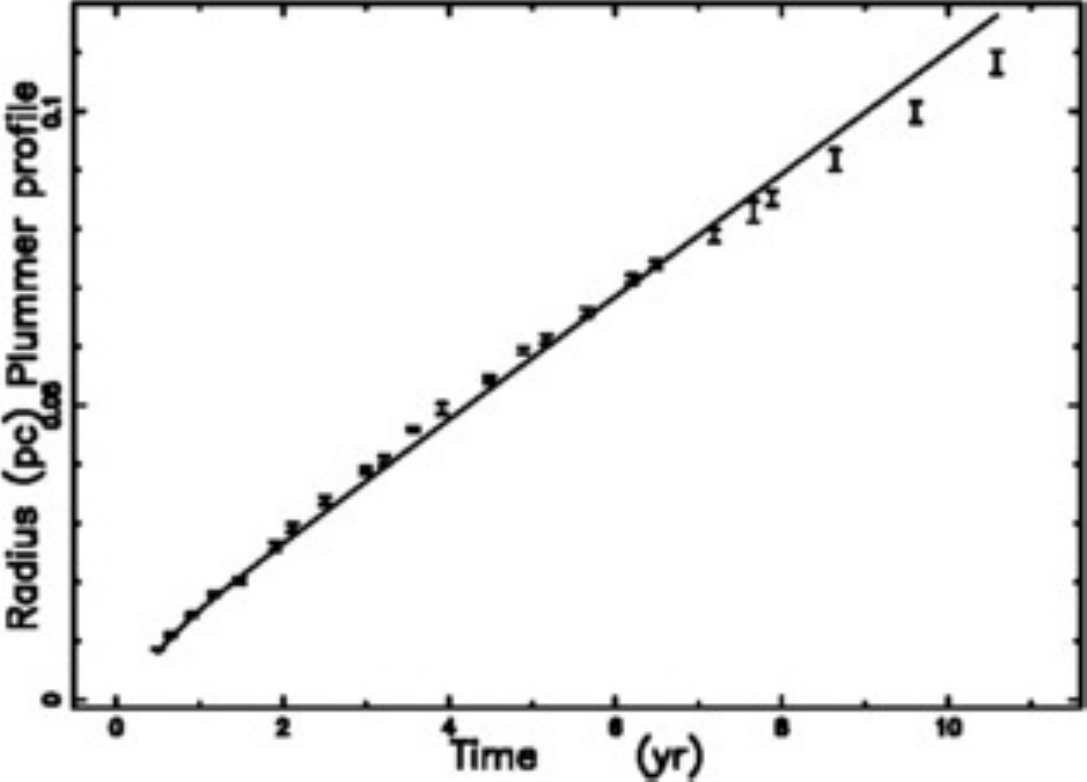}
\end {center}
\caption
{
Theoretical radius for the Plummer-type profile as obtained
by the  solution of  the nonlinear
equation connected with (\ref{eqndiffplummer})
(full line).
Data as in Table \ref{datafit}.
The astronomical data of \snr is represented with
vertical error bars.
}
\label{snr_1993_plummer}
    \end{figure*}

The numerical solution of the differential equation
connected with the Plummer-like profile, $\eta=6$,
is shown in  Figure~\ref{snr_1993_plummer} when
the data of Table \ref{datafit} is adopted.

\begin{figure*}
\begin{center}
\includegraphics[width=7cm ]{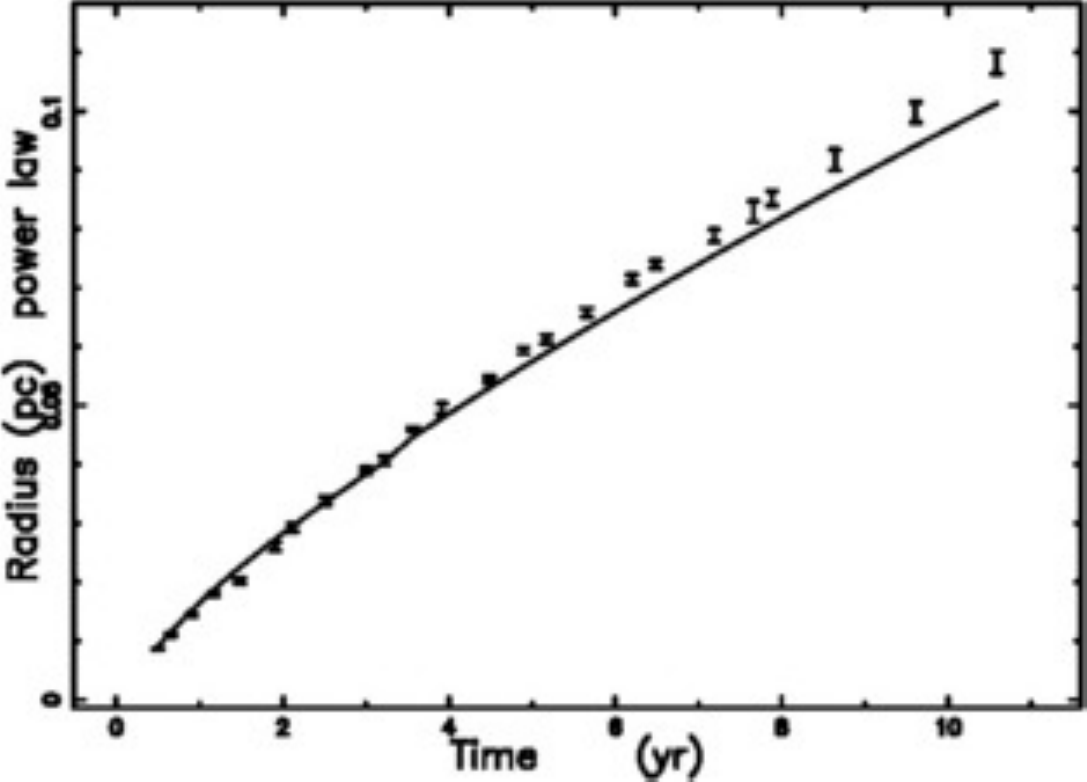}
\end {center}
\caption
{
Theoretical radius for the power law case as obtained
by the  solution of  the nonlinear
equation connected with (\ref{eqndiffpowerlaw})
(full line).
Data as in Table \ref{datafit}.
The astronomical data of \snr is represented with
vertical error bars.
}
\label{snr_1993_emden_power}
    \end{figure*}

A comparison with the power law behavior for the CSM
is shown in Figure~\ref{snr_1993_emden_power},
which is built from the data in Table \ref{datafit}.

The series solution for the power law dependence of the CSM
with coefficients as given by  equation~(\ref{coefficientspower})
is not reported  because
the  range in time
of its reliability is limited to $t-t_0 \approx 0.0003$\ yr.

\section{Relativistic  applications to \snr}

\label {applicationsrelativistic}

We now apply the relativistic solutions  derived so far to \snr
in the  Lane--Emden and Plummer cases.

\subsection{Relativistic Lane--Emden case ($n=5$)}

\begin{figure*}
\begin{center}
\includegraphics[width=7cm ]{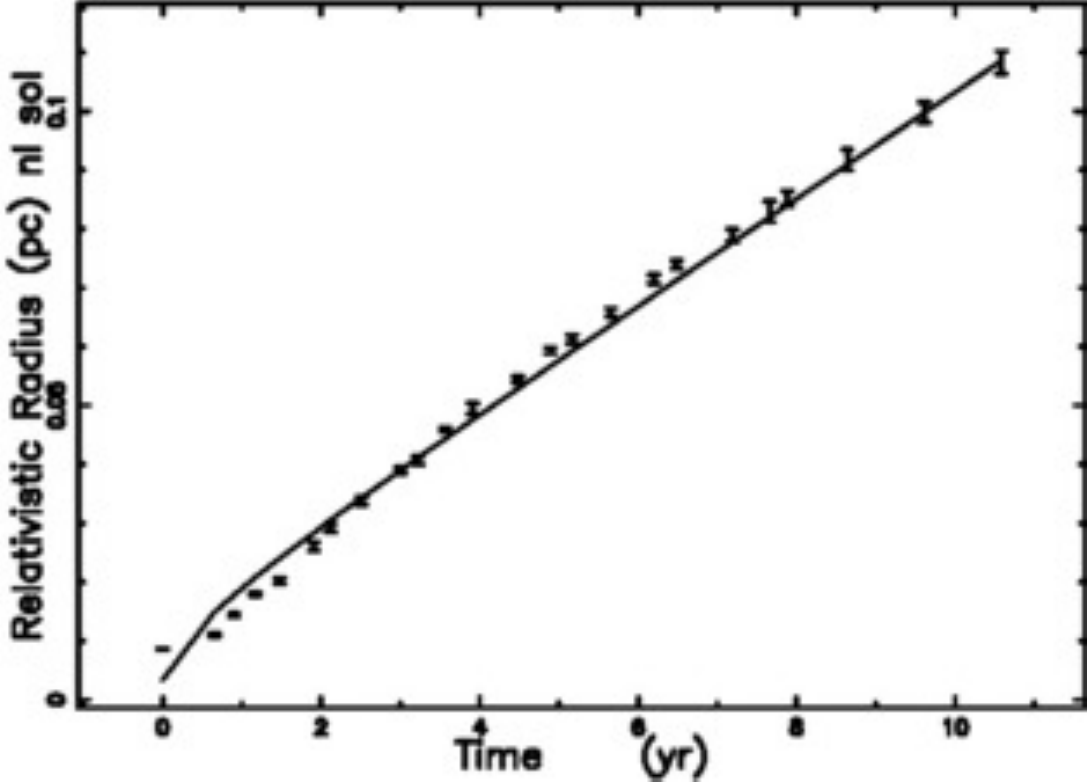}
\end {center}
\caption
{
Theoretical relativistic radius as solution of the
nonlinear equation (\ref{nlequation}) (full line),
with data as in Table~\ref{datafitrel}:
Lane--Emden case ($n=5$).
The  astronomical data of \snr
is represented with vertical error bars.
\label{sol_analytical}
}
    \end{figure*}

\begin{figure*}
\begin{center}
\includegraphics[width=7cm ]{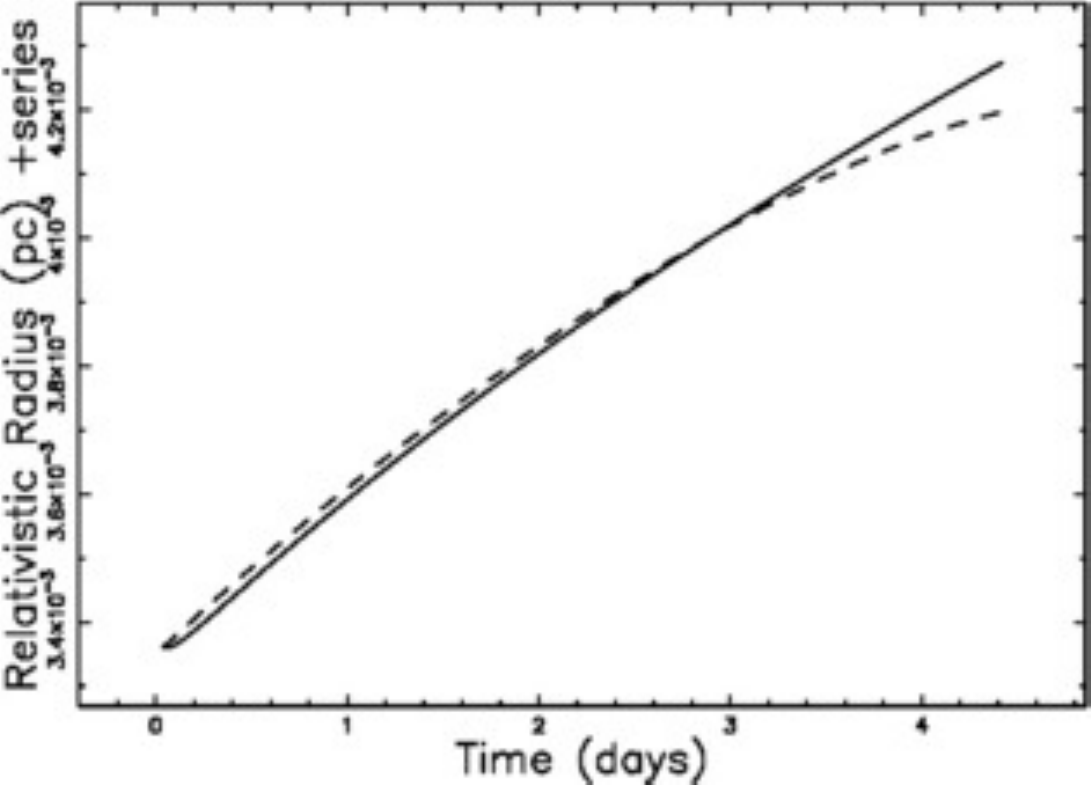}
\end {center}
\caption
{
Theoretical relativistic radius as solution of the
nonlinear equation (\ref{nlequation}) (full line),
and series solution as
given by equation~(\ref{rtseriesrel}) (dashed line):
Lane--Emden case ($n=5$)
Data as in Table~\ref{datafitrel}.
The time is expressed in days.
}
\label{sol_series}
    \end{figure*}

\begin{figure*}
\begin{center}
\includegraphics[width=7cm ]{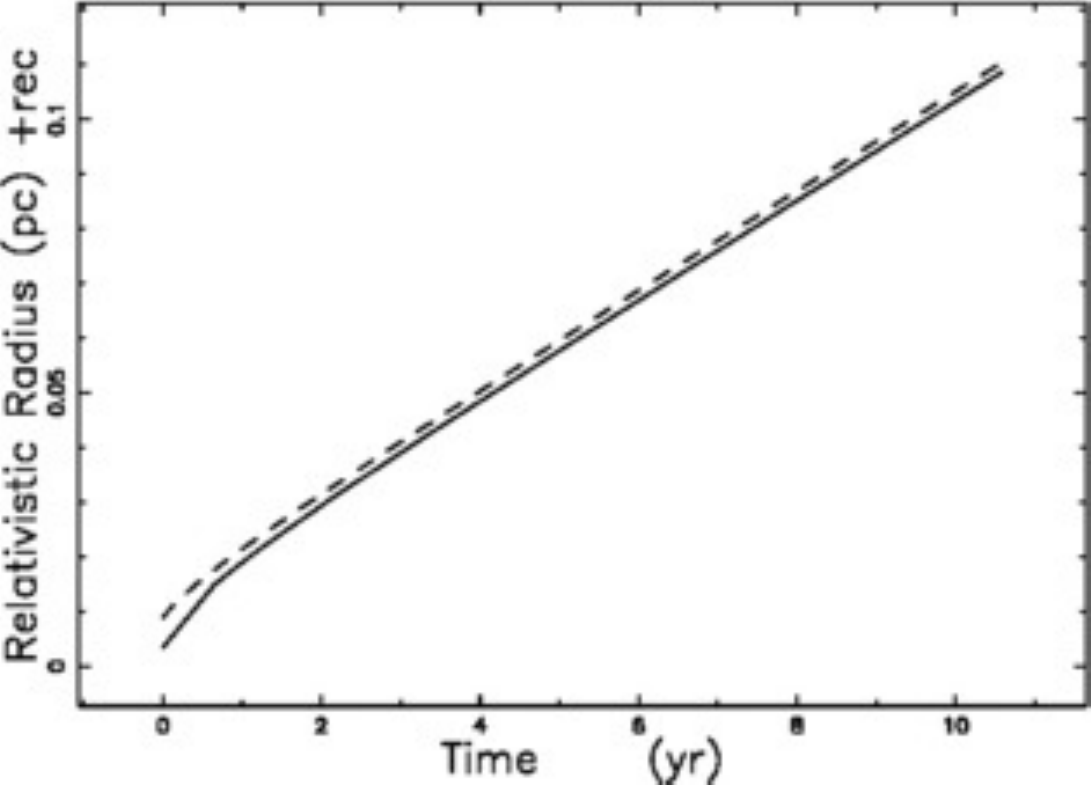}
\end {center}
\caption
{
Theoretical relativistic radius as solution of the
nonlinear equation (\ref{nlequation}) (full line),
and recursive solution as
given by equation~(\ref{recursiverel})
when $\Delta t=0.053 \mathrm{yr}$ (dashed line):
Lane--Emden case ($n=5$)
Data as in Table~\ref{datafitrel}.
}
\label{sol_recursive}
    \end{figure*}

The initial observed velocity, $v_0$,  as deduced
from  radio observations,
see \cite{Marcaide2009}, is $v_0 \approx 20000\,\velu$  at
$t_0 \approx 0.5 \, \mathrm{ yr}$.
We now reduce the initial time $t_0$ and we increase the velocity
up to the relativistic regime, $t_0=10^{-4}\, \mathrm{yr}$,
and $v_0= 100000\,\velu$.
This choice of parameters allows fitting
the observed radius--time
relation that  should be reproduced.
The  data used in the simulation is
shown in Table \ref{datafitrel}.
\begin{table}
\caption
{
Numerical values of the parameters
used in  three relativistic solutions
for the Lane--Emden case ($n=5$)
}
 \label{datafitrel}
 \[
 \begin{array}{c}
 \hline
 \hline
 \noalign{\smallskip}
 parameters      \\
  t_0=10^{-4}~\mathrm{yr}~;
  r_0=0.0033~\mathrm{pc} ~;
~\beta_0=0.3333 ~;
~ b=0.004~ \mathrm{pc}
\\
\noalign{\smallskip} \hline
 \end{array}
 \]
 \end {table}
The relativistic numerical solution of equation~(\ref{nlequation})
is shown in Figure~\ref{sol_analytical},
the relativistic series solution as given
by (\ref{rtseriesrel}) is shown in Figure~\ref{sol_series},
and Figure~\ref{sol_recursive} contains the
recursive solution  as given by equation~(\ref{recursiverel}).

%
%
%

\subsection{Relativistic Plummer case ($\eta=6$)}

\begin{figure*}
\begin{center}
\includegraphics[width=7cm ]{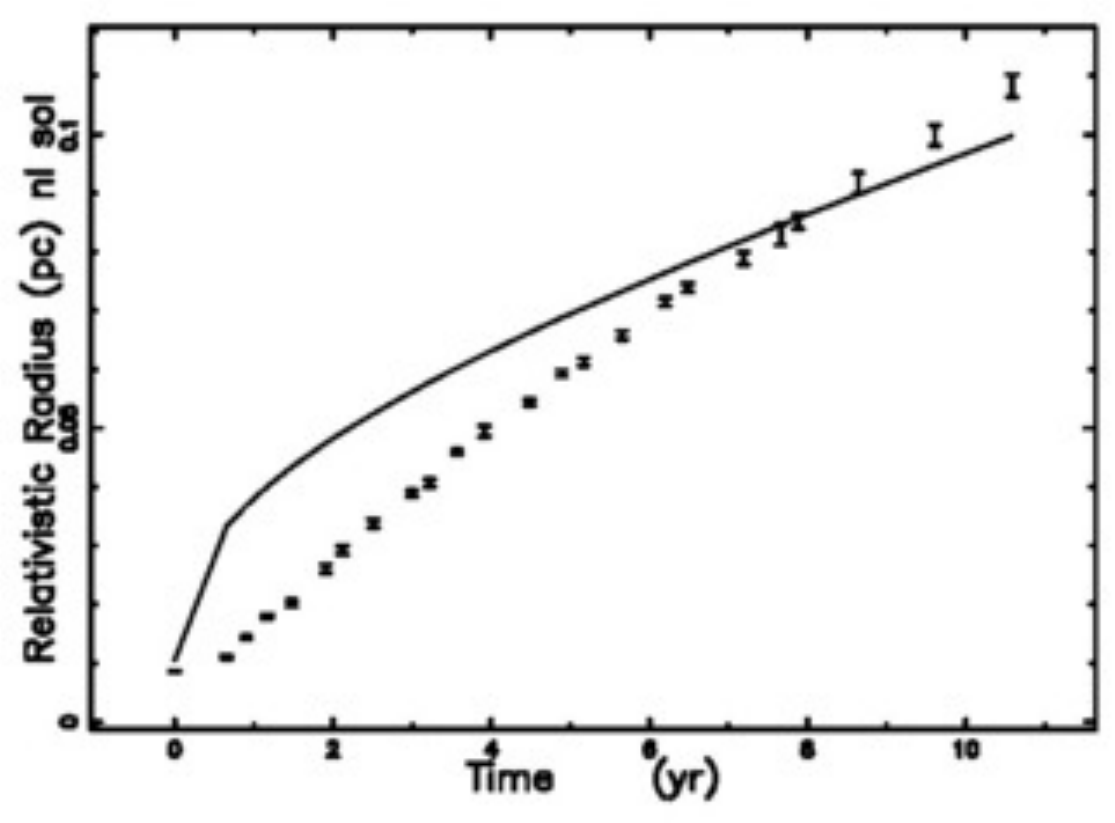}
\end {center}
\caption
{
Theoretical relativistic radius as solution of the
nonlinear equation (\ref{eqndiffrelplummer}) (full line),
with data as in Table~\ref{datafitrelplummer}
and astronomical data of \snr with
vertical error bars:
Plummer  case ($\eta=6$)
\label{relatistic_radius_years}
}
    \end{figure*}

\begin{figure*}
\begin{center}
\includegraphics[width=7cm ]{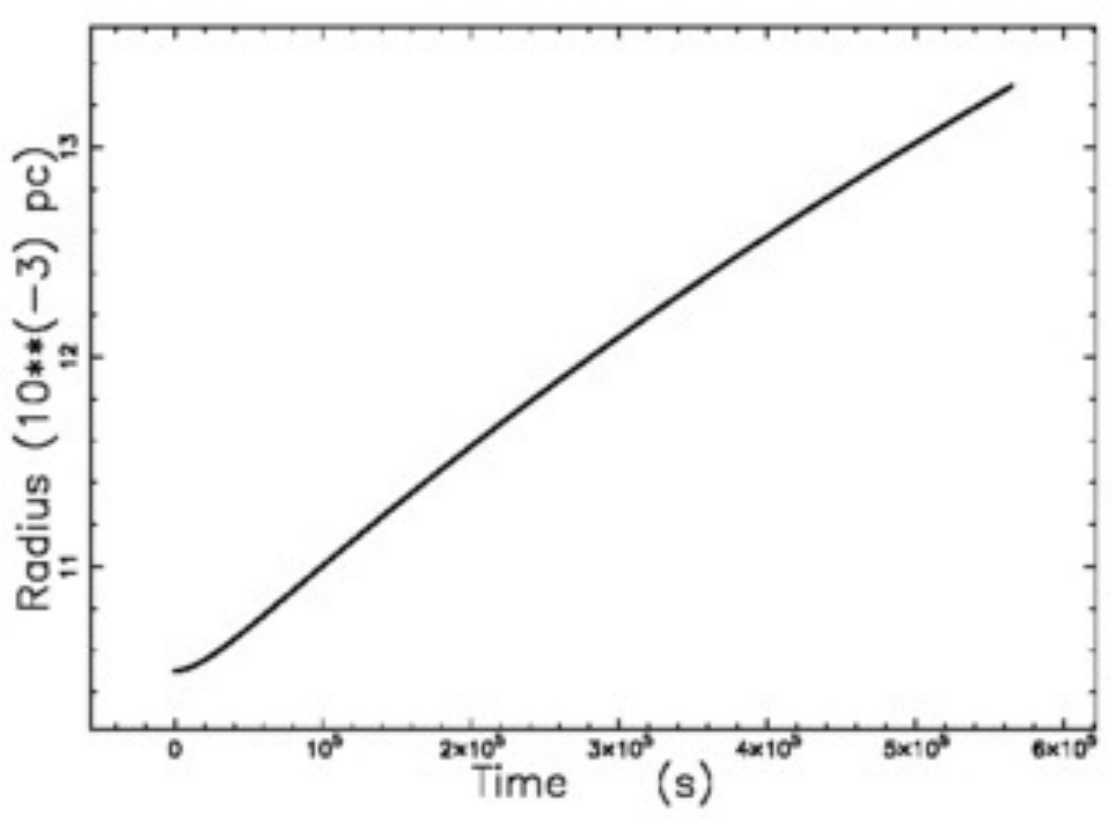}
\end {center}
\caption
{
Theoretical relativistic radius 
in units of $10^{-3}$ pc 
as solution of the
nonlinear equation (\ref{eqndiffrelplummer}) (full line),
with data as in Table~\ref{datafitrelplummer}
and time in seconds:
Plummer  case ($\eta=6$) 
\label{relatistic_radius_seconds}
}
    \end{figure*}

The relativistic numerical solution of equation 
(\ref{eqndiffrelplummer})
is shown in Figure~\ref{relatistic_radius_years}
with parameters as in Table~\ref{datafitrelplummer}.
Figure~\ref{relatistic_radius_seconds} shows the
theoretical radius in units of $10^{-3}$\ pc
as a function of the time in seconds.

\begin{table}
\caption
{
Numerical values of the parameters
used in the Plummer ($\eta=6$)  
relativistic solution.
}
 \label{datafitrelplummer}
 \[
 \begin{array}{c}
 \hline
 \hline
 \noalign{\smallskip}
 parameters      \\
  t_0=1 \times 10^{-7}~\mathrm{yr}~or~t_0=3.15~\mathrm{seconds}~\\
  R_0=0.01~\mathrm{pc} \\
 \beta_0=0.666 \\
  b=0.028~ \mathrm{pc}\\
  d=1 \\
\noalign{\smallskip} \hline
 \end{array}
 \]
 \end {table}

%

%

\section{Classical  Applications to SNR}
\label{secastrosnr}

\begin{figure*}
\begin{center}
\includegraphics[width=7cm ]{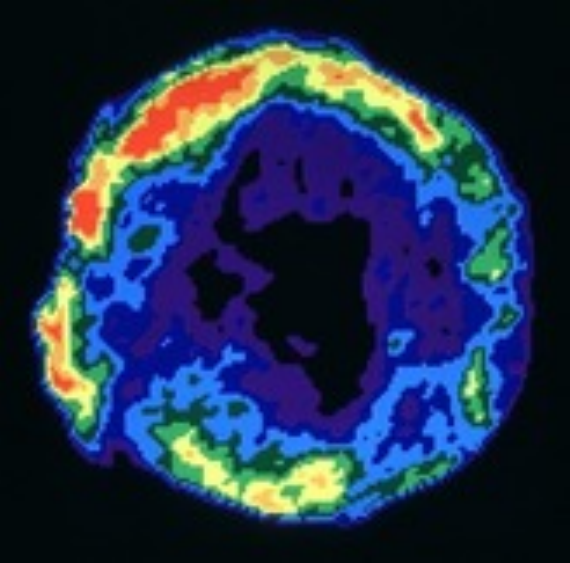}
\end {center}
\caption
{
Radio image  of 
Tycho or 3C10.
}
\label{Tycho}
    \end{figure*}

\begin{figure*}
\begin{center}
\includegraphics[width=7cm ]{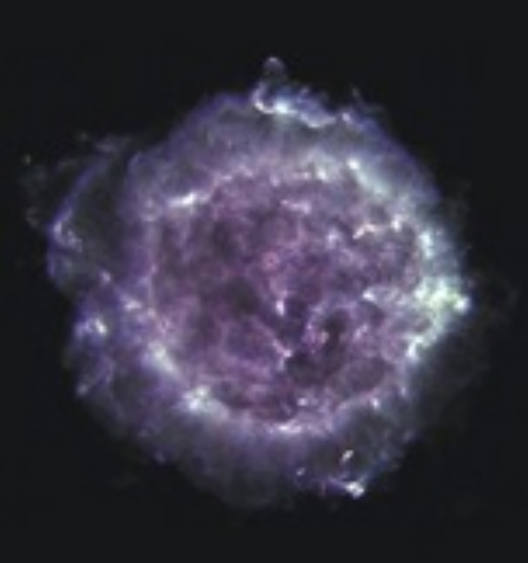}
\end {center}
\caption
{
Image  of 
CasA at 3 different frequencies: 
1.4 GHz (L band), 5.0 GHz (C band), and 8.4
GHz (X band).
}
\label{CasA}
    \end{figure*}

\begin{figure*}
\begin{center}
\includegraphics[width=7cm ]{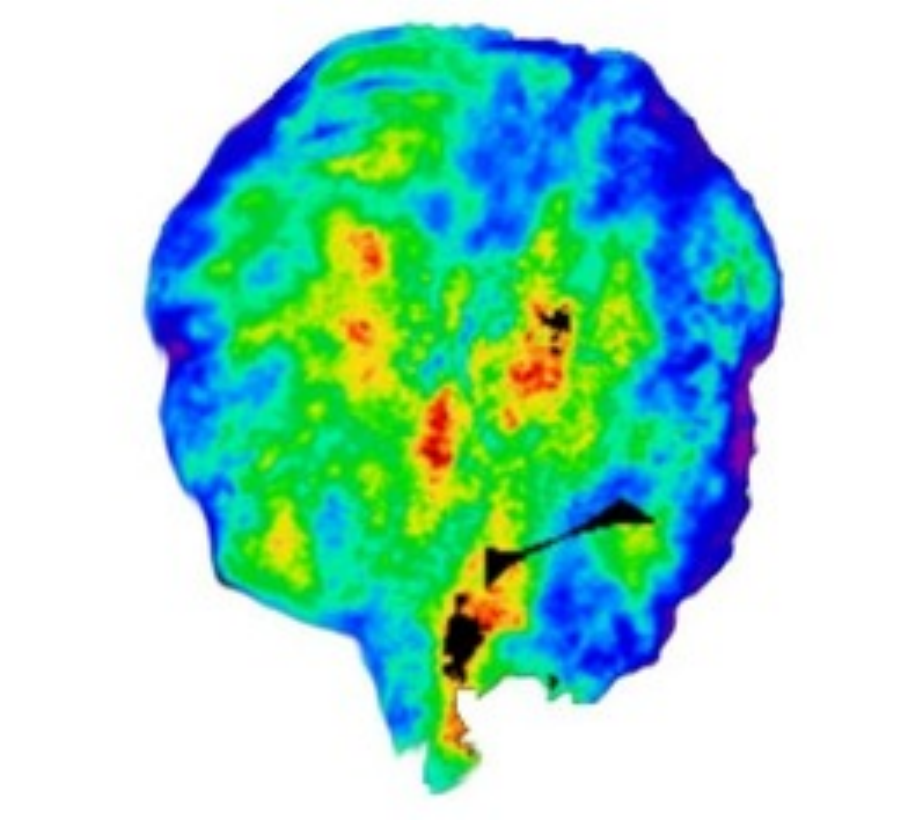}
\end {center}
\caption
{
X image  of the 
Cygnusloop.
}
\label{Cygnusloop}
    \end{figure*}

\begin{figure*}
\begin{center}
\includegraphics[width=7cm ]{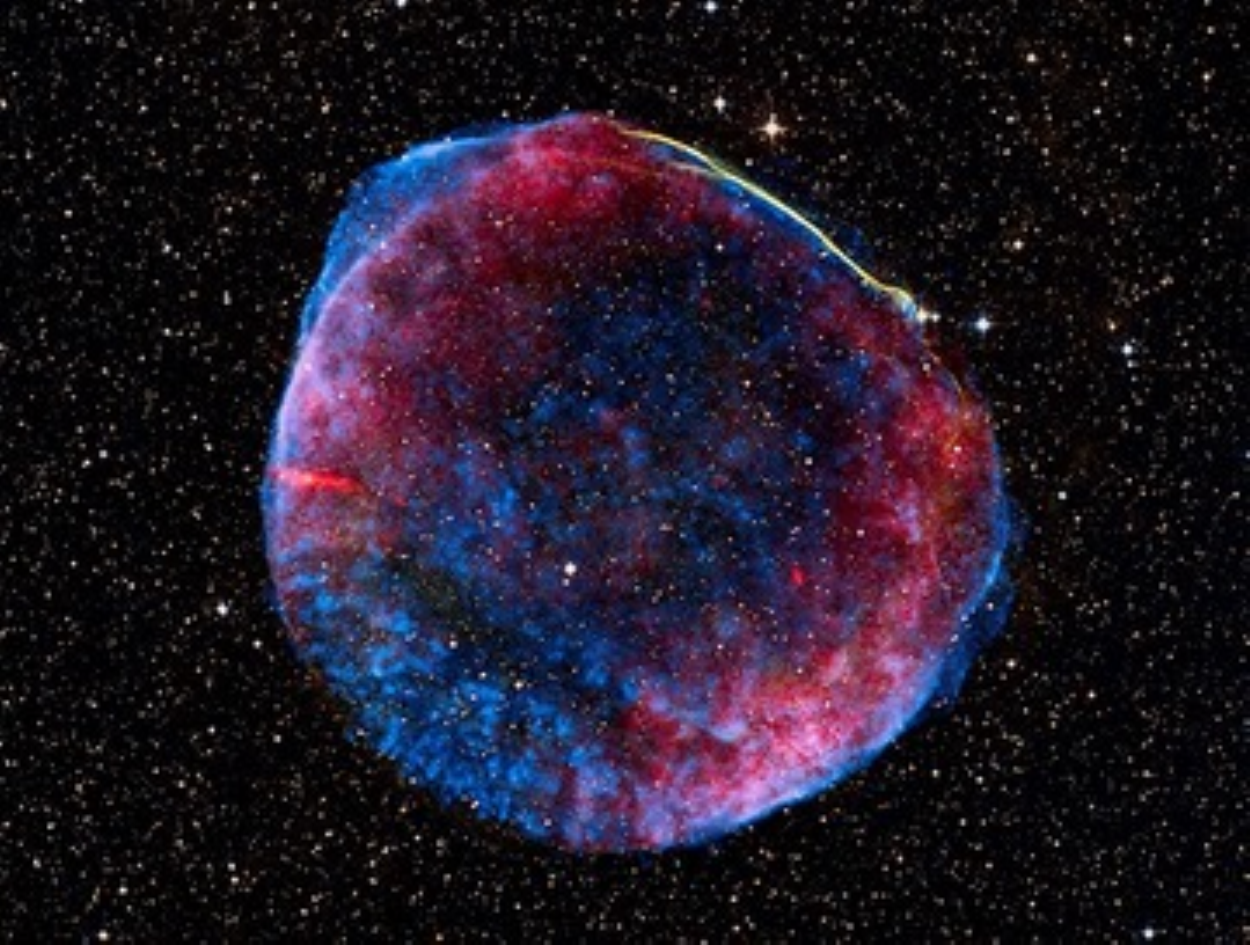}
\end {center}
\caption
{
Composite image of SN~1006: 
X-ray data in blue, optical data in yellowish hues, 
and radio image data in red. 
}
\label{sn1006}
    \end{figure*}

In the previous section, we derived three equations of motion 
in the form of nonlinear  equations 
and three Pad\'e approximated equations of motion.
We now check the reliability of the numerical and approximate 
solutions on four SNRs: Tycho or 3C10, see \cite{Williams2016}
and the radio map  in  
Figure \ref{Tycho},
Cas A, see \cite{Patnaude2009}
and the radio map  in Figure \ref{CasA},  
Cygnus loop,  see \cite{Chiad2015}
and the X-map in \ref{Cygnusloop},
and  SN~1006, see \cite{Uchida2013}
and  image in \ref{sn1006}.

The three astronomical measurable parameters 
are the time since the explosion in years, $t$,
the actual observed radius in pc, $r$,
and the present velocity of expansion in 
km\,s$^{-1}$, see Table \ref{tablesnrs}.
\begin{table}[ht!]
\caption {
Observed astronomical parameters of SNRs
}
\label{tablesnrs}
\begin{center}
\begin{tabular}{|c|c|c|c|c|}
\hline
Name  &  Age (yr)   &  Radius (pc) & Velocity (km\,s$^{-1}$)& References\\
\hline                                                   
Tycho        & 442        &  3.7         & 5300 & Williams~et~al.~2016    \\
Cas ~A       & 328        &  2.5         & 4700 & Patnaude~and~Fesen~2009 \\
Cygnus~loop  & 17000      &  24.25       & 250  & Chiad~et~al.~2015       \\
SN ~1006     & 1000       &  10.19       & 3100 & Uchida~et~al.2013       \\
\hline
\end{tabular}
\end{center}
\end{table}
The astrophysical  units have not yet been specified:
pc for length  and  yr for time
are the units most commonly used by astronomers.
With these units, the initial velocity is 
$v_0(km s^{-1})= 9.7968 \, 10^5 v_0(pc\,yr^{-1})$.
The determination of the four  unknown parameters, which are   
$t_0$, $r_0$, $v_0$ and $b$,  
can be obtained by equating the observed astronomical velocities 
and radius with those obtained with the 
Pad\'e rational polynomial, i.e.,
\begin{eqnarray}
\label{eqnnl1}
        r_{2,1}&= Radius(pc),\\
\label{eqnnl2}
        v_{2,1}&=Velocity(km s^{-1}). 
\end{eqnarray}

In order to reduce the unknown parameters from four to two, we
fix $v_0$ and  $t_0$.
The two parameters $b$ and $r_0$ are found by solving the
two nonlinear equations (\ref{eqnnl1})  
and (\ref{eqnnl2}).
The results for the three types of profiles here adopted 
are shown in Tables 
\ref{tablesnrsexp},
\ref{tablesnrsgauss}
and
\ref{tablesnrslaneemden}.

\begin{table}[ht!]
\caption {
Theoretical  parameters of SNRs
for the Pad\'e  approximated equation of motion 
with an exponential profile. 
}
\label{tablesnrsexp}
\begin{center}
\begin{tabular}{|c|c|c|c|c|c|c|}
\hline
Name         &$t_0$(yr)&$r_0$(pc)&$v_0(km\,s^{-1})$& b(pc)&$\delta\,(\%)$&
$\Delta\,v (km\,s^{-1}) $ \\
\hline                                                   
Tycho        &  0.1  &  1.203    & 8000   &  0.113  &   5.893  & -1.35   \\
Cas ~A       &  1    &  0.819    & 8000   &  0.1    &   6.668  & -3.29   \\
Cygnus~loop  &  10   &  12.27    & 3000   & 45.79   &   6.12   & -0.155  \\
SN ~1006     &  1    &  5.49     & 10000  & 2.332   &   1.455  & -12.34  \\
\hline
\end{tabular}
\end{center}
\end{table}

\begin{table}[ht!]
\caption {
Theoretical  parameters of SNRs
for the Pad\'e  approximated equation of motion 
with a Gaussian  profile. 
}
\label{tablesnrsgauss}
\begin{center}
\begin{tabular}{|c|c|c|c|c|c|c|}
\hline
Name         &$t_0$(yr)&$r_0$(pc)&$v_0(km\,s^{-1})$& b(pc)&$\delta\,(\%)$&
$\Delta\,v (km\,s^{-1}) $ \\
\hline                                                   
Tycho        &  0.1  &  1.022    & 8000   &  0.561   &   8.517  & -10.469   \\
Cas ~A       &  1    &  0.741    & 7000   &  0.406   &   7.571  & -13.16   \\
Cygnus~loop  &  10   &  11.92    & 3000   &  21.803  &   7.875  & -0.161  \\
SN ~1006     &  1    &  5.049    & 10000  &  4.311   &   4.568  & -18.58  \\
\hline
\end{tabular}
\end{center}
\end{table}

\begin{table}[ht!]
\caption 
{
Theoretical  parameters of SNRs
for the Pad\'e  approximated equation of motion 
with a  Lane--Emden profile. 
}
\label{tablesnrslaneemden}
\begin{center}
\begin{tabular}{|c|c|c|c|c|c|c|}
\hline
Name         &$t_0$(yr)&$r_0$(pc)&$v_0(km\,s^{-1})$& b(pc)&$\delta\,(\%)$&
$\Delta\,v (km\,s^{-1}) $ \\
\hline                                                   
Tycho        &  0.1  &   0.971    & 8000   & 0.502   &    3.27   & -14.83   \\
Cas ~A       &  1    &   0.635    & 8000   & 0.35    &    4.769  & -23.454   \\
Cygnus~loop  &  10   &   11.91    & 3000   & 27.203  &    7.731  & -0.162  \\
SN ~1006     &  1    &   5        & 10000  & 4.85    &    3.297  & -19.334   \\
\hline
\end{tabular}
\end{center}
\end{table}
The    goodness of the approximation is evaluated
through the percentage error, $\delta$, which is
\begin{equation}
\delta = \frac{\big | r_{2,1} - r_E \big |}
{r_E} \times 100
\quad ,
\end{equation}
where $r_{2,1}$ is the 
Pad\'e approximated radius and
$r_E$ is the exact solution which is obtained by solving numerically
the nonlinear equation of motion, as an example equation~(\ref{eqn_nl_exp})
in the exponential case.
The numerical values of $\delta$ are shown 
in  column 6 of  
Tables 
\ref{tablesnrsexp},
\ref{tablesnrsgauss}
and
\ref{tablesnrslaneemden}.
Another useful astrophysical variable  is the 
predicted decrease in velocity 
on the basis of the     
Pad\'e approximated velocity, $v_{2,1}$, over ten years,
see column 7 of  
Tables 
\ref{tablesnrsexp},
\ref{tablesnrsgauss}
and
\ref{tablesnrslaneemden}.
More details can be found  in \cite{Zaninetti2017b}.

\section{Conclusions}

{\bf Classic Case: Lane--Emden type }

The thin layer approximation  which models
the expansion  in a self-gravitating medium
of the Lane--Emden type ($n=5$)
can be modeled
by a differential  equation of the first order
for the radius as a function of  time.
This  differential  equation has
an analytical solution represented by equation~(\ref{radiusemdent}).
A power law series, see equation~(\ref{rtseries}),
can model the solution
of the Lane--Emden type
for a limited range of time,
see Figure~\ref{1993pc_fit_series} relative to \snr.
A  recursive solution for the first
order differential equation,
as represented by equation~(\ref{recursive}), approximates
quite well the analytical solution of the Lane--Emden type
 and at the time
of $t=10$ yr a precision of four digits is reached
when $\Delta t=10^{-3}$ yr.
The Pad\'e rational polynomial 
approximation 
allows deriving an approximate solution 
for the advancing radius, see equation~(\ref{rmotionlaneemden}),  
and for the velocity, see equation~(\ref{vmotionlaneemden}).

{\bf Classic Case: Plummer case }

The   differential  equation 
which represents the expansion in the Plummer case ($\eta=6$) 
is equation~\ref{eqndiffplummer}: 
no analytical   solution exists, 
but a numerical solution is shown in  
Figure~\ref{snr_1993_plummer} 
relative to \snr.

{\bf Relativistic case: Lane--Emden case}

The temporal  evolution of an SN in a self-gravitating medium
of the Lane--Emden type can be found by applying the
conservation of relativistic momentum in the thin layer
approximation.
This  relativistic invariant is  determined by
a differential equation of the first order, see
equation~(\ref{eqndiffrel}).
Three different relativistic solutions   for the radius as a function
of time  have been derived:
(i) a numerical  solution,
see equation~(\ref{nlequation})
which covers the range $10^{-4}\mathrm{ yr} <; \, t <; 10\, \mathrm{yr} $
and
fits the observed radius--time relation
for \snr;
(ii) a series solution, see equation~(\ref{nlequation}),
which has a limited
range of validity,
$10^{-4}\mathrm{ yr} <; \, t <; 1.2\times10^{-2} \, \mathrm{yr} $;
(iii) a recursive solution 
in which the  desired accuracy  is  reached
by  decreasing the time step $\Delta t$.
The relativistic results here presented
model \snr and   are obtained with
an  initial velocity of $v_0=100 000\velu$ or $\beta_0=0.333$
or $\gamma=1.06$.

{\bf Relativistic case: Plummer  case}

The temporal  evolution of an SN in a  medium
of the Plummer  type, $\eta=6$,   can be found by applying the
conservation of relativistic momentum in the thin layer
approximation.
This  relativistic invariant is  found via 
a differential equation of the first order, see
equation~(\ref{eqndiffrelplummer}).
A series solution  is shown in equation~(\ref{rtseriesrelplummer6}) relative to \snr.

{\bf Classic SNRs}

The application of the 
Pad\'e  approximant to the left-hand side of the complicated 
equation of motion 
allows finding three approximate laws of motion,
see equations~(\ref{rmotionexp}, \ref{rmotiongauss}, and \ref{rmotionlaneemden}),
and three approximate velocities, 
see equations~(\ref{vmotionexp}, \ref{vmotiongauss}, and \ref{vmotionlaneemden}). 
An astrophysical test was performed 
on four SNRs assumed to be spherical;
the four sets  of  parameters  were 
presented in  Tables  
\ref{tablesnrsexp},
\ref{tablesnrsgauss}
and
\ref{tablesnrslaneemden}.
The percentage of error of the Pad\'e  approximated 
solutions for the radius is always less than 
$10\%$ with respect to the numerical exact solution,
see column 6 of  
the three last tables.
In order to produce an astrophysical prediction,
the theoretical decrease in velocity for the four SNRs here
analysed has been evaluated,
see column 7 of  
Tables 
\ref{tablesnrsexp},
\ref{tablesnrsgauss}
and
\ref{tablesnrslaneemden}.

\section*{Acknowledgments}

Credit for Figures 
\ref{SN1993J},
\ref{Tycho},
\ref{CasA}
 is  given to  
 the National Radio Astronomy Observatory Image Gallery.
Credit for Figure \ref{Cygnusloop}  is given 
to the High Energy Astrophysics Science Archive Research Center.
Credit for Figure \ref{sn1006}  is given 
to Astronomy Picture of the Day.


\begin{thebibliography}{10}
\expandafter\ifx\csname url\endcsname\relax
  \def\url#1{\texttt{#1}}\fi
\expandafter\ifx\csname urlprefix\endcsname\relax\def\urlprefix{URL }\fi

\bibitem{Wang2003}
{Wang}, L., {Baade}, D., {Hoflich}, P., {Khokhlov}, A., \& {Wheeler}, J.~C.,
  {Spectropolarimetry of SN 2001el in NGC 1448: Asphericity of a Normal Type Ia
  Supernova}, \apj 591 (2003) 1110.

\bibitem{Mazzali2005}
{Mazzali}, P.~A., {Benetti}, S., {Altavilla}, G., {Blanc}, G., \& {Cappellaro},
  E., {High-Velocity Features: A Ubiquitous Property of Type Ia Supernovae},
  \apjl 623 (2005) L37.

\bibitem{Marion2013}
{Marion}, G.~H., {Vinko}, J., \& {Wheeler}, J.~C., {High-velocity Line Forming
  Regions in the Type Ia Supernova 2009ig}, \apj 777 (2013) 40.

\bibitem{Childress2014}
{Childress}, M.~J., {Filippenko}, A.~V., {Ganeshalingam}, M., \& { Schmidt},
  B.~P., {High velocity features in Type Ia supernova spectra}, \mnras 437
  (2014) 338.

\bibitem{Branch2017}
{Branch}, D. \& {Wheeler}, J.~C., {Supernova Explosions}, Springer-Verlag,
  Berlin, 2017.

\bibitem{Silverman2015}
{Silverman}, J.~M., {Vink{\'o}}, J., {Marion}, G.~H., {et~al.}, {High-velocity
  features of calcium and silicon in the spectra of Type Ia supernovae}, \mnras
  451 (2015) 1973.

\bibitem{Marcaide2009}
{Marcaide}, J.~M., {Mart{\'{\i}}-Vidal}, I., {Alberdi}, A., \&
  {P{\'e}rez-Torres}, M.~A., {A decade of SN 1993J: discovery of radio
  wavelength effects in the expansion rate}, \aap 505 (2009) 927.

\bibitem{Sedov1959}
{Sedov}, L.~I., {Similarity and Dimensional Methods in Mechanics}, Academic
  Press, New York, 1959.

\bibitem{Dalgarno1987}
{McCray}, R.~A., {Coronal interstellar gas and supernova remnants}, in:
  {A.~Dalgarno \& D.~Layzer} (Ed.), Spectroscopy of Astrophysical Plasmas,
  {Cambridge University Press. }, Cambridge, 1987, 255--278.

\bibitem{Dyson1997}
{{Dyson}, J.~E. and {Williams}, D.~A.}, {The physics of the interstellar
  medium}, Institute of Physics Publishing, Bristol, 1997.

\bibitem{Padmanabhan_II_2001}
{Padmanabhan}, P., {Theoretical astrophysics. Vol. II: Stars and Stellar
  Systems}, {Cambridge University Press}, {Cambridge, UK}, {2001}.

\bibitem{Zaninetti2011a}
{Zaninetti}, L., {Time-dependent models for a decade of SN 1993J}, \apss 333
  (2011) 99.

\bibitem{Zaninetti2012c}
Zaninetti, L., An adjustable law of motion for relativistic spherical shells,
  Central European Journal of Physics 10 (2012) 32.

\bibitem{Chevalier1982a}
{Chevalier}, R.~A., {Self-similar solutions for the interaction of stellar
  ejecta with an external medium}, \apj 258 (1982) 790.

\bibitem{Chevalier1982b}
{Chevalier}, R.~A., {The radio and X-ray emission from type II supernovae},
  \apj 259 (1982) 302.

\bibitem{Kompaneets1960}
{Kompaneyets}, A.~S., {A Point Explosion in an Inhomogeneous Atmosphere },
  Soviet Phys. Dokl. 5 (1960) 46.

\bibitem{Olano2009}
{Olano}, C.~A., {The propagation of the shock wave from a strong explosion in a
  plane-parallel stratified medium: the Kompaneets approximation}, \aap 506
  (2009) 1215.

\bibitem{Granot2006}
{Granot}, J. \& {Kumar}, P., {Distribution of gamma-ray burst ejecta energy
  with Lorentz factor}, \mnras 366 (2006) L13.

\bibitem{Peer2007}
{Pe'er}, A., {Ryde}, F., {Wijers}, R.~A.~M.~J., \& {M{\'e}sz{\'a}ros}, P., {A
  New Method of Determining the Initial Size and Lorentz Factor of Gamma-Ray
  Burst Fireballs Using a Thermal Emission Component}, \apjl 664 (2007) L1.

\bibitem{Zou2010}
{Zou}, Y.-C. \& {Piran}, T., {Lorentz factor constraint from the very early
  external shock of the gamma-ray burst ejecta}, \mnras 402 (2010) 1854.

\bibitem{Aoi2010}
{Aoi}, J., {Murase}, K., {Takahashi}, K., {Ioka}, K., \& {Nagataki}, S., {Can
  We Probe the Lorentz Factor of Gamma-ray Bursts from GeV-TeV Spectra
  Integrated Over Internal Shocks?}, \apj 722 (2010) 440.

\bibitem{Muccino2013}
{Muccino}, M., {Ruffini}, R., {Bianco}, C.~L., \& {Izzo}, L., {GRB 090510: A
  Disguised Short Gamma-Ray Burst with the Highest Lorentz Factor and
  Circumburst Medium}, \apj 772 (2013) 62.

\bibitem{Spitzer1978}
{Spitzer}, L., {Physical processes in the interstellar medium}, Wiley,
  New-York, 1978.

\bibitem{NIST2010}
Olver, F. W. J.~e., Lozier, D. W.~e., Boisvert, R. F.~e., \& Clark, C. W.~e.,
  {NIST handbook of mathematical functions.}, {Cambridge University Press. },
  Cambridge, 2010.

\bibitem{Plummer1911}
{Plummer}, H.~C., {On the problem of distribution in globular star clusters},
  \mnras 71 (1911) 460.

\bibitem{Whitworth2001}
{Whitworth}, A.~P. \& {Ward-Thompson}, D., {An Empirical Model for Protostellar
  Collapse}, \apj 547 (2001) 317.

\bibitem{Abramowitz1965}
{Abramowitz}, M. \& {Stegun}, I.~A., {Handbook of Mathematical Functions with
  Formulas, Graphs, and Mathematical Tables}, Dover, New York, 1965.

\bibitem{Zaninetti2014f}
{Zaninetti}, L., { A classical and a relativistic law of motion for spherical
  supernovae }, \apj 795 (2014) 80.

\bibitem{Lane1870}
{Lane}, H.~J., On the theoretical temperature of the sun, under the hypothesis
  of a gaseous mass maintaining its volume by its internal heat, and depending
  on the laws of gases as known to terrestrial experiment, American Journal of
  Science 148 (1870) 57.

\bibitem{Emden1907}
{Emden}, R., Gaskugeln: anwendungen der mechanischen w{a}rmetheorie auf
  kosmologische und meteorologische probleme, B. Teubner., Berlin, 1907.

\bibitem{Chandrasekhar_1967}
{Chandrasekhar}, S., {An introduction to the study of stellar structure}, New
  York, 1967.

\bibitem{Binney2011}
{Binney}, J. \& {Tremaine}, S., {Galactic dynamics, Second Edition}, Princeton
  University Press, Princeton, NJ, 2011.

\bibitem{Zwillinger1989}
{Zwillinger}, D., {Handbook of differential equations}, Academic Press, New
  York, 1989.

\bibitem{Hansen1994}
{Hansen}, C.~J. \& {Kawaler}, S.~D., {Stellar Interiors. Physical Principles,
  Structure, and Evolution.}, Springer-Verlag, Berlin, 1994.

\bibitem{Tenenbaum1963}
{Tenenbaum}, M. \& {Pollard}, H., Ordinary Differential Equations: An
  Elementary Textbook for Students of Mathematics, Engineering, and the
  Sciences, Dover Publications, New York, 1963.

\bibitem{Ince2012}
{Ince}, E.~L., Ordinary differential equations, Courier Dover Publications, New
  York, 2012.

\bibitem{French1968}
{{French}, A.P.}, {Special Relativity}, {CRC}, New~York, 1968.

\bibitem{Zhang1997}
{Zhang}, Y., Special Relativity and Its Experimental Foundations, World
  Scientific, Singapore, 1997.

\bibitem{Guery2010}
Gu{\'e}ry-Odelin, D. \& Lahaye, T., Classical Mechanics Illustrated by Modern
  Physics: 42 Problems with Solutions, Imperial College Press, London, 2010.

\bibitem{Larmor1897}
{Larmor}, J., {A dynamical theory of the electric and luminiferous medium. Part
  III. Relations with material media.}, {Philos. Trans. R. Soc. Lond., Ser. A,
  Contain. Pap. Math. Phys. Character} 190 (1897) 205.

\bibitem{Lorentz1904}
Lorentz, H.~A., Electromagnetic phenomena in a system moving with any velocity
  smaller than that of light, in: Proc. Acad. Sciences Amsterdam, Vol.~6, 1904,
  809--830.

\bibitem{Einstein1905}
{Einstein}, A., {Zur Elektrodynamik bewegter K{o}rper}, Annalen der Physik 322
  (1905) 891.

\bibitem{Macrossan1986}
{Macrossan}, M.~N., A note on relativity before einstein, The British journal
  for the philosophy of science 37~(2) (1986) 232.

\bibitem{Bevington2003}
{{Bevington}, P.~R. and {Robinson}, D.~K.}, {Data Reduction and Error Analysis
  for the Physical Sciences}, McGraw-Hill, New York, 2003.

\bibitem{Bartel2007}
{Bartel}, N., {Bietenholz}, M.~F., {Rupen}, M.~P., \& {Dwarkadas}, V.~V., {SN
  1993J VLBI. IV. A Geometric Distance to M81 with the Expanding Shock Front
  Method}, \apj 668 (2007) 924.

\bibitem{Aldering1994}
{Aldering}, G., {Humphreys}, R.~M., \& {Richmond}, M., {SN 1993J: The optical
  properties of its progenitor}, \aj 107 (1994) 662.

\bibitem{Maund2004}
{Maund}, J.~R., {Smartt}, S.~J., {Kudritzki}, R.~P., {Podsiadlowski}, P., \&
  {Gilmore}, G.~F., {The massive binary companion star to the progenitor of
  supernova 1993J}, \nat 427 (2004) 129.

\bibitem{Weaver1977}
{Weaver}, R., {McCray}, R., {Castor}, J., {Shapiro}, P., \& {Moore}, R.,
  {Interstellar bubbles. II - Structure and evolution}, \apj 218 (1977) 377.

\bibitem{Schmidt1993}
{Schmidt}, B.~P., {Kirshner}, R.~P., {Eastman}, R.~G., {et~al.}, {The unusual
  supernova SN1993J in the galaxy M81}, \nat 364 (1993) 600.

\bibitem{Smith2008a}
{Smith}, N., {Galactic Twins of the Ring Nebula Around SN1987A and a Possible
  LBV-like Phase for Sk-69 202}, in: Revista Mexicana de Astronomia y
  Astrofisica Conference Series, Vol.~33 of Revista Mexicana de Astronomia y
  Astrofisica Conference Series, 2008, 154--156.

\bibitem{Williams2016}
{Williams}, B.~J., {Chomiuk}, L., {Hewitt}, J.~W., {et~al.}, {An X-Ray and
  Radio Study of the Varying Expansion Velocities in Tycho Supernova Remnant},
  \apjl 823 (2016) L32.

\bibitem{Patnaude2009}
{Patnaude}, D.~J. \& {Fesen}, R.~A., {Proper Motions and Brightness Variations
  of Nonthermal X-ray Filaments in the Cassiopeia A Supernova Remnant}, \apj
  697 (2009) 535.

\bibitem{Chiad2015}
{Chiad}, B.~T., {Ali}, L.~T., \& {Hassani}, A.~S., {Determination of Velocity
  and Radius of Supernova Remnant after 1000 yrs of Explosion}, International
  Journal of Astronomy and Astrophysics 5 (2015) 125.

\bibitem{Uchida2013}
{Uchida}, H., {Yamaguchi}, H., \& {Koyama}, K., {Asymmetric Ejecta Distribution
  in SN 1006}, \apj 771 (2013) 56.

\bibitem{Zaninetti2017b}
{Zaninetti}, L., Pad\'e approximant for the equation of motion of a supernova
  remnant, Journal of High Energy Physics, Gravitation and Cosmology 3 (2017)
  78.

\end{thebibliography}

\label{lastpage-01}
\end{document}